\documentclass[10pt]{iopart}
\usepackage{graphicx}
\eqnobysec

\newcommand{\beq}{\begin{equation}}
\newcommand{\beqa}{\begin{eqnarray}}
\newcommand{\eeq}{\end{equation}}
\newcommand{\eeqa}{\end{eqnarray}}
\newcommand{\cotanh}{\mathop{\rm cotanh}\nolimits}
\newcommand{\cum}[1]{{\overline{#1}}^{({\rm c})}}
\renewcommand{\d}{{\rm d}}
\newcommand{\ds}{\displaystyle}
\newcommand{\eps}{\varepsilon}
\newcommand{\erf}{\mathop{\rm erf}\nolimits}
\newcommand{\frad}[2]{\ds{\frac{\ds#1}{\ds#2}}}
\newcommand{\ii}{{\rm i}}
\renewcommand{\max}{{\rm max}}
\renewcommand{\min}{{\rm min}}
\newcommand{\mean}[1]{\langle#1\rangle}
\newcommand{\prob}[1]{\mathop{\rm Prob}\nolimits\{#1\}}
\newcommand{\qbar}{{/\!\!\!q}}

\newcommand{\sg}{{\rm sg}}
\newcommand{\var}{\mathop{\rm var}\nolimits}
\newcommand{\C}{{\bf C}}
\newcommand{\GC}{^{{\rm GC}}}
\newcommand{\I}{{\cal I}}
\newcommand{\Z}{{\cal Z}}

\begin{document}

\title
[Condensation in the ZRP:
an interplay between interaction and disorder]
{Condensation in the inhomogeneous zero-range process:
an interplay between interaction and diffusion disorder}

\author{C Godr\`eche and J M Luck}

\address{Institut de Physique Th\'eorique, CEA Saclay
and URA 2306, CNRS, 91191~Gif-sur-Yvette cedex, France}

\begin{abstract}
We analyze the role of the interplay between on-site interaction and inhomogeneous
diffusion on the phenomenon of condensation in the zero-range process.
We predict a universal phase diagram in the plane of two exponents,
respectively characterizing the interactions and the diffusion disorder.
The most prominent outcome is the existence of an extended condensed phase.
In the latter phase,
which originates as a result of the combined effects of strong enough interaction
and weak enough disorder,
a typical high-density configuration
has a unique condensate on top of a critical background,
but the condensate may be located at any site of a large hosting set of favored sites,
whose size grows sub-extensively.
The novel extended condensed phase thus interpolates continuously
between the two scenarios associated so far with the condensation transition,
namely spontaneous symmetry breaking and explicit symmetry breaking.
\end{abstract}

\eads{\mailto{claude.godreche@cea.fr},\mailto{jean-marc.luck@cea.fr}}

\maketitle

\section{Introduction}
\label{intro}

The essence of the zero-range process (ZRP) can be best understood by returning
to the original work of Spitzer~\cite{spitz}.
The starting point is to consider an assembly of independent particles diffusing
on a graph, i.e., a finite connected set of $M$ sites.
The particles, which we can also refer to as random walkers,
are also in finite number,~$N$.
Choosing a continuous-time description of the process, any individual particle
hops from site $m$ to site $n$ with a rate $w_{mn}$.
In the course of time, the probabilities for the particle to be at the various sites $m$
relax to the unique normalized solutions $q_m$
of the stationary master equation
\beq
\sum_{n=1}^M w_{nm}q_n=q_m\sum_{n=1}^M w_{mn}.
\label{master}
\eeq

For the assembly of independent random walkers, if at time $t$ the occupation
of site~$m$ is equal to $k$, then, of course, the chance of seeing a particle
hop from this site to site~$n$ is enhanced by a factor $k$, which amounts to
multiplying the hopping rate $w_{mn}$ by~$k$.
The intention stated by Spitzer was to depart from the field of independent
random walkers by introducing a form of interaction between them.
In Spitzer's original definition of the zero-range process,
the rate $k\,w_{mn}$ for the hopping of a particle from site $m$,
with occupation equal to $k$, to site $n$,
which holds if the particles are independent,
is changed to the product
\beq
W(m,n,k)=w_{mn}\,u_k,
\label{zrp}
\eeq
where the factor $u_k$ only depends on the occupation $k$ of the
departure site~$m$ at the current time.
This dependence gives an explicit content to the denomination of the
process as `zero-range'.
This very definition of the ZRP demonstrates the importance and respective
roles of the two facets of this process:
{\it diffusion} on the one hand, {\it interaction} on the other hand.
The first one is encoded in the single-particle hopping rate $w_{mn}$,
while the second one is encoded in the on-site interaction factor~$u_k$.
Either factor of the product~(\ref{zrp})
can be separately normalized in an arbitrary way~(see below~(\ref{pk})).
This justifies that either factor will sometimes be referred to as a `rate'.

The interest of the statistical physics community for the ZRP came
long after the publication of~\cite{spitz} and the subsequent works of
mathematicians on the subject~\cite{andj,land}.
The popularity of the ZRP was triggered by the ubiquity of the phenomenon of
condensation in a number of processes related in one way or another to the ZRP
(for reviews, see~\cite{braz,hanney,lux}).
The condensation transition manifests itself by the existence
of a critical density $\rho_c$, with the following meaning.
Consider the stationary state of the model
in the thermodynamic limit where $M$ and $N\to\infty$,
at fixed density $\rho=N/M$.
If the density is high enough ($\rho>\rho_c$),
a macroscopic (extensive) number of excess particles, of order
\beq
\Delta=N-M\rho_c=M(\rho-\rho_c),
\label{ddef}
\eeq
concentrates on a single site with very high probability.

In the present work we analyze the role of the interplay between the two
facets of the ZRP, diffusion and interaction,
on the phenomenon of condensation.
More precisely, we consider the ZRP defined by the rate~(\ref{zrp})
on a finite system of $N$ particles living on $M$ sites,
with the following specifications.

\begin{itemize}

\item
{\it Diffusion disorder.}
The single-particle weights $q_m$ given by~(\ref{master})
enter the factorized form~(\ref{pfac})
of the stationary distribution of the occupations.
Rather than explicitly modeling diffusion disorder by endowing
a given structure with random hopping rates~$w_{mn}$
and calculating the single-particle weights $q_m$ through~(\ref{master}),
we instead assume that the $q_m$
can be modeled as {\it independent random variables} drawn from a given distribution.
We will expand on the motivation for this assumption in section~\ref{ourmodel}.
More precisely,
we characterize the inhomogeneity of the diffusion process
by modeling the $q_m$ as i.i.d.~random variables
drawn from the continuous distribution
\beq
f(q)=c(1-q)^{c-1}\qquad(0<q<1),
\label{fdef}
\eeq
with a power-law singularity as $q\to1$ with an arbitrary exponent $c>0$,
to be referred to as the {\it disorder exponent}.
The shape of the distribution $f(q)$
in the vicinity of the maximum (supposed to be finite and set here to $q_\max=1$)
will play a central role in the analysis of the condensation transition.

\item
{\it Interaction.}
We choose for the factor $u_k$, which encodes the interaction between particles
sitting at the same site, the form
\beq
u_k=1+\frac{b}{k},
\label{rate}
\eeq
where $b$ is a parameter describing the strength of on-site interactions,
to be referred to as the {\it interaction exponent}.

\end{itemize}

The above model contains as limiting cases two known models exhibiting a condensation phenomenon
(see references below).

\begin{itemize}

\item[(i)] {\it Homogeneous ZRP.}
If diffusion is homogeneous
($q_m=1$ for all $m$, i.e., $f(q)=\delta(q-1)$, or formally $c=0$),
the process with rate~(\ref{rate}) is known to have a condensation transition
at the critical density
\beq
\rho_c=\frac{1}{b-2},
\label{rhoc1}
\eeq
whenever the interactions are sufficiently attractive ($b>2$).
In this situation, the condensate can sit on any site with equal probabilities
in the stationary state of the system.
This condensation scenario therefore corresponds to a
{\it spontaneous symmetry breaking} (SSB) of the symmetry between the sites.

\item[(ii)] {\it Occupation-independent inhomogeneous ZRP.}
If the rate $u_k$ is occupation-independent ($u_k=1$, i.e., $b=0$),
the process with inhomogeneous diffusion described by the distribution~(\ref{fdef})
is also known to have a condensation transition, at the critical density
\beq
\rho_c=\frac{1}{c-1},
\label{rhoc2}
\eeq
whenever the disorder is sufficiently strong ($c>1$).
In this situation
the condensate is typically located on the site with the largest
single-particle weight.
This corresponds to a scenario of {\it explicit symmetry breaking} (ESB).

\end{itemize}

The full model defined by diffusion disorder~(\ref{fdef})
and interaction~(\ref{rate})
interpolates between the two limiting cases recalled above.
The aim of this work is to investigate to what extent
the condensation transition of the homogeneous ZRP defined in (i)
is altered by the presence of inhomogeneous diffusion,
i.e., of the non-trivial distribution~(\ref{fdef}) of the single-particle weights $q_m$,
or, conversely, how the condensation transition of the inhomogeneous ZRP defined in (ii)
is altered by the presence of an interaction
corresponding to the rate~(\ref{rate}).

\begin{figure}[!ht]
\begin{center}
\includegraphics[angle=0,width=.5\linewidth]{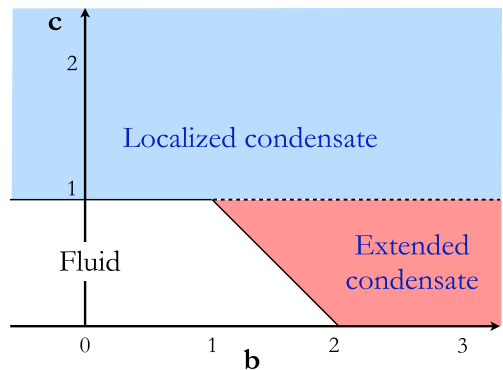}
\caption{\small
Universal phase diagram of the inhomogeneous ZRP in the $b$--$c$ plane,
with its fluid phase and its two condensed phases:
a localized one ($c>1$) and a novel extended one ($c<1$).}
\label{phase1}
\end{center}
\end{figure}

\begin{figure}[!ht]
\begin{center}
\includegraphics[angle=0,width=.5\linewidth]{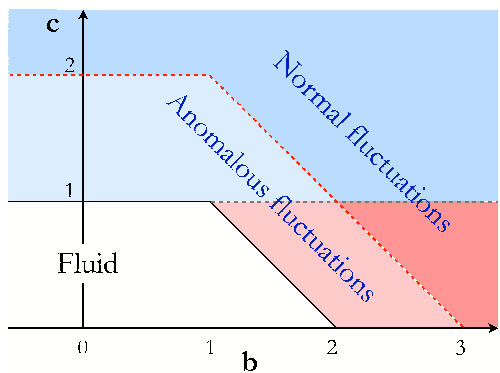}
\caption{\small
More detailed phase diagram, showing the internal structure of the condensed phases.
The size fluctuations of the condensate around its mean $\Delta$
are either normal (finite mean square occupation),
or anomalous (divergent mean square occupation).}
\label{phase2}
\end{center}
\end{figure}

An informal account of our results is as follows.
We show in figure~\ref{phase1}
the phase diagram of our model in the plane of the
exponents $b$ (characterizing interaction)
and~$c$ (characterizing disorder).
In the {\it fluid phase}, there is no condensate at any density.
In the {\it localized condensed phase},
for any $\rho>\rho_c$, a condensate of size $\Delta$ (see~(\ref{ddef}))
lives on the site with the largest
single-particle weight with very high probability.
In the {\it extended condensed phase},
for any $\rho>\rho_c$, the condensate is still unique with very high probability,
but it may be located at any site of a large {\it hosting set} of favored sites,
whose size~$R$ grows sub-extensively~as
\beq
R\sim\frac{M}{\Delta^c}\sim\frac{M^{1-c}}{(\rho-\rho_c)^c}.
\eeq
The existence of an extended condensed phase is the most salient novel feature
put forward in the present work.
The latter phase
implements a continuous interpolation
between the two symmetry breaking scenarios recalled above,
namely SSB in the homogeneous ZRP ($c=0$, $b>2$),
and ESB in the occupation-independent inhomogeneous ZRP ($b=0$, $c>1$).
These two limiting scenarios are consistently recovered as $R=M$ and $R=1$, respectively.
The ESB scheme more generally applies to the ZRP with $c>1$ and any $b$.
Figure~\ref{phase2} shows a more refined phase diagram,
where the condensed phases are further divided into a regime of {\it normal fluctuations},
where the mean square occupation~$\mu_c$ at criticality is finite,
so that the size fluctuations of the condensate around its mean size $\Delta$
are Gaussian and grow as $M^{1/2}$,
and a regime of {\it anomalous fluctuations}, where $\mu_c$ is divergent,
so that size fluctuations are not Gaussian and grow faster than $M^{1/2}$.
The phase diagrams depicted in figures~\ref{phase1} and~\ref{phase2} are universal,
in the sense that they hold
for any rate whose asymptotic behavior at large $k$
is of the form $u_k\approx1+b/k$,
and for any distribution of single-particle weights
exhibiting a finite maximum $q_\max$
and a power-law singularity $f(q)\sim(q_\max-q)^{c-1}$ near this maximum.

Before proceeding, let us review the relevant literature.

The homogeneous ZRP with rate~(\ref{rate}) has been studied in a long series of
works, and is by now well understood~\cite{braz,hanney,lux}.
The main features of the statics of the model without reference to a
dynamical process were first analyzed in~\cite{bia}\footnote{The equilibrium state of the static model considered in~\cite{bia} is actually slightly different from the stationary state of the homogeneous ZRP with rate~(\ref{rate}).
However the universal static properties of the two models are the same.
The same holds for the dynamics, i.e., the universal properties of the dynamics of the homogeneous ZRP with rate~(\ref{rate}) or of the process considered in~\cite{zeta1,zeta2} are the same.
This was underlined in~\cite{cg}.}.
The underlying dynamical process was introduced in~\cite{zeta1}, where the
coarsening dynamics leading to condensation was studied in the mean-field
geometry of the complete graph.
This study was subsequently pursued in~\cite{zeta2}.
The relationship of the equilibrium state of the model described in~\cite{bia}
to the stationary state of a ZRP was recognized in~\cite{loan} and a first
analysis of the criterion on the form of the rate $u_k$ for condensation to
occur was given.
This analysis was subsequently completed in~\cite{braz}.
Independently, a similar analysis was performed in~\cite{jeon}, devoted to
a study of the size of the condensate.
Further studies involve the analysis of the coarsening process leading to
condensation in the one-dimensional case~\cite{cg,gross}, or on the complete
graph~\cite{cg}, the proof of the uniqueness of the condensate and the equivalence
of ensembles~\cite{gross}, the analysis of the motion of the condensate at
stationarity~\cite{us}, the analysis of finite-size effects in the canonical
ensemble~\cite{maj}, and the relationship of the condensation phenomenon
to extreme-value statistics~\cite{maj3}.
More mathematical works on similar subjects can be found in~\cite{armendariz,fls,landim}.

The occupation-independent ZRP with rate $u_k=1$
and inhomogeneous diffusion described by~(\ref{fdef}) can be traced back
to studies of one-dimensional disordered totally asymmetric exclusion
models~\cite{bfl,krug96,evans96}, where the connections with the ZRP~\cite{bfl,krug96}
and the relationship to Bose-Einstein condensation~\cite{evans96} were recognized.
A more mathematical treatment of the condensation transition
has been given in~\cite{afgl},
while the dynamics of the same model was analyzed in~\cite{jain},
both at stationarity and during the process of the formation of the condensate.

The references cited so far led to the current understanding
of the two limiting prototypical cases of the ZRP referred to above.
We now turn to other works which deal in some way or another
with the interplay of interaction and disorder in the ZRP.
The simplest situation of an inhomogeneous ZRP
is that of an attractive impurity in a homogeneous background:
one site is singled out by its weight $q>1$,
while all the other sites keep $q=1$.
In this situation condensation occurs even for the occupation-independent ZRP
($u_k=1$)~\cite{braz,angel}.
The totally asymmetric one-dimensional inhomogeneous ZRP
has been investigated from a rigorous standpoint in~\cite{landim1}.
The situation where the diffusion disorder originates
in the structure of the underlying graph has also been analyzed~\cite{noh,krakow,tlz},
the main emphasis being on complex (scale-free) networks.
In~\cite{noh} complete condensation, i.e., $\rho_c=0$,
has been shown to generally occur on such networks.
References~\cite{krakow} also deal with the statics and the dynamics
of the occupation-independent ZRP on complex networks.
The phenomenon of condensation in disordered urn models has been explored in~\cite{ohkubo}.
The more general case of an inhomogeneous ZRP where disorder enters the interaction
part of the rate itself has been recently considered in~\cite{chleboun}.
Finally, the occupation-independent one-dimensional ZRP with partially asymmetric
disordered diffusion has been analyzed in~\cite{jsi}.
Reference~\cite{barma} presents a brief review of the more general area
of driven diffusive systems with disorder.

The setup of the present paper is as follows.
In section~\ref{generalities}, we present some general formalism
and emphasize two special situations, namely $u_k=k$
(independent random walkers) and $u_k=1$ (the occupation-independent case,
which can be mapped onto an ideal Bose gas).
We then review, in section~\ref{limits},
the two limiting cases of interest recalled above,
namely the homogeneous ZRP with rate~(\ref{rate})
and the occupation-independent inhomogeneous ZRP.
Section~\ref{thermo} contains our theoretical investigations,
whose main predictions are summarized in the universal phase diagrams
depicted in figures~\ref{phase1} and~\ref{phase2}.
Our quantitative predictions are then illustrated by means
of numerical results in section~\ref{numerics}.
Section~\ref{networks} presents an application of our findings
to the case of complex networks.
Section~\ref{discussion} contains a brief discussion,
whereas more technical aspects are exposed in three appendices:
\ref{sectwosites} contains an investigation of the ZRP on two sites,
while~\ref{borderline} is devoted to the universal fluctuations
of the hosting probabilities in the borderline situation ($c=1$),
and~\ref{effloc}
to an effective model describing the relevant rare events in the localized regime ($c>1$).

\section{General formalism}
\label{generalities}

\subsection{Stationary state}

We start by a reminder of the definition of the ZRP and of the
properties of its stationary state.
Consider a finite connected graph of~$M$ sites (labelled $m=1,\dots,M$),
on which particles hop from site to site in continuous time.
We denote by $N_m$ the instantaneous occupation of site~$m$,
i.e., the number of particles living at site $m$ at the current time.
The total number of particles in the system,
\beq
N=\sum_{m=1}^M N_m,
\label{ntot}
\eeq
is conserved by the dynamics.

According to Spitzer's original work~\cite{spitz},
a ZRP is defined by the following expression (see~(\ref{zrp}))
\beq
W(m,n,k)=w_{mn}\,u_k
\label{zrp1}
\eeq
for the rate for a particle to hop from site $m$ (with occupation $N_m=k$) to site $n$,
where $w_{mn}$ describes the diffusion of a single particle on the graph,
and $u_k$ encodes the interaction between particles sitting at the same site.
In the present work we focus our attention on an inhomogeneous ZRP defined by~(\ref{zrp1}),
i.e., with diffusion disorder.
For completeness, we mention the following more general definition of a ZRP~\cite{hanney},
\beq
W(m,n,N_m)=w_{mn}\,u_{m,k},
\label{zrp2}
\eeq
where the factor $u_{m,k}$ encoding the interaction between the $k$ particles
sitting at site~$m$ depends explicitly on both $m$ and $k$.
References~\cite{chleboun} contain an investigation
of such a situation, where quenched disorder affects the
interactions between particles, and not only the diffusion.

The fundamental property of the ZRP is the simple structure of its stationary state:
the distribution of the occupations is factorized.
Such a distribution is referred to as a product measure.
The particles being considered as indistinguishable,
a configuration of the system is entirely defined by the occupations~$\{N_m\}$.
The stationary probability $P(\{N_m\})$ of any configuration
is given by a product of elementary weights,
associated with the occupations of each site.
For the hopping rate~(\ref{zrp}) (or~(\ref{zrp1})), we have
\beq
P(\{N_m\})=\frac{1}{Z_{M,N}}\,\prod_{m=1}^M
q_m^{N_m}\,p_{N_m}\delta\!\left(\sum_{m=1}^M N_m,N\right).
\label{pfac}
\eeq
The Kronecker delta function ensures the condition~(\ref{ntot}),
whereas the normalisation factor $Z_{M,N}$ (see~(\ref{zdef})) plays the role of
a partition function,
and the weight associated with each site~$m$ is itself a product of the two factors,
$q_m^{N_m}$ and $p_{N_m}$.
The latter factor is given by
\beq
p_0=1,\qquad p_k=\frac{1}{u_1\dots u_k}.
\label{pk}
\eeq

As long as one is only interested in static properties,
either factor of the rate~(\ref{zrp}) (or~(\ref{zrp1}))
can be normalized in an arbitrary way.
Indeed, if all the hopping rates~$w_{mn}$ are multiplied by the same constant~$A$,
the time scale of the dynamical process is changed as $t\to t/A$,
but the single-particle weights $q_m$ obeying the stationary master equation~(\ref{master})
are left unchanged.
Similarly, if all the factors $u_k$ are multiplied by the same constant $B$,
$p_k$ is changed to $p_k/B^k$,
the partition function~$Z_{M,N}$ gets multiplied by $1/B^N$,
and so the result~(\ref{pfac}) is unchanged.
Furthermore, the single-particle weights themselves
can also be normalized in an arbitrary way.
Indeed, if all the~$q_m$ are multiplied by the same constant $C$,
$Z_{M,N}$ gets multiplied by $C^N$,
and so the result~(\ref{pfac}) is again unchanged.

\subsubsection*{Canonical ensemble.}

In the canonical ensemble, the total number $N$ of particles is fixed.
The stationary probability $P(\{N_m\})$ of any configuration is given by the
expression~(\ref{pfac}) above, where the partition function
\beq
Z_{M,N}=\sum_{\{N_m\}}\prod_{m=1}^M q_m^{N_m}\,p_{N_m}
\,\delta\!\left(\sum_{m=1}^M N_m,N\right)
\label{zdef}
\eeq
depends on all the $q_m$ and all the $p_k$.

Singling out the sum over $N_m=k$ in~(\ref{zdef})
leads to the recursion relation
\beq
Z_{M,N}=\sum_{k=0}^N q_m^k\,p_k\,Z_{M-1,N-k}(\qbar_m)
\label{zrec}
\eeq
between partition functions of systems with successive sizes.
The argument $\qbar_m$ means that $Z_{M-1,N-k}(\qbar_m)$
depend on the $(M-1)$ single-particle weights $q_n$ for $n\ne m$.
Likewise the distribution of the occupation of site $m$ is obtained
from~(\ref{pfac}) as
\beq
f_{m,k}=\prob{N_m=k}=q_m^k\,p_k\,\frac{Z_{M-1,N-k}(\qbar_m)}{Z_{M,N}}.
\label{fres}
\eeq
Hence the mean occupation (local density) at site $m$ reads
\beq
\rho_m=\mean{N_m}=\sum_{k=0}^N kf_{m,k}
=\frac{q_m}{Z_{M,N}}\,\frac{\partial Z_{M,N}}{\partial q_m}
=q_m\frac{\partial}{\partial q_m}\ln Z_{M,N}.
\label{rhores}
\eeq
The partition function $Z_{M,N}$ being a homogeneous function
of degree $N$ of all the~$q_m$, the sum rule
\beq
\sum_{m=1}^M\rho_m=N
\eeq
is ensured by Euler's identity for homogeneous functions:
\beq
\sum_{m=1}^Mq_m\,\frac{\partial Z_{M,N}}{\partial q_m}=NZ_{M,N}.
\eeq

By using a contour-integral representation of the Kronecker delta function,
we can recast the partition function~(\ref{zdef}) into the form
\beq
Z_{M,N}=\oint\frac{\d z}{2\pi\ii z^{N+1}}\,\Z_M(z),
\label{contour}
\eeq
where the grand partition function reads
\beq
\Z_M(z)=\sum_{N\ge0}Z_{M,N}z^{N}=\prod_{m=1}^M P(zq_m),
\eeq
and
\beq
P(z)=\sum_{k\ge0}p_kz^k.
\eeq

\subsubsection*{Grand-canonical ensemble.}

The representation~(\ref{contour}) of the partition function
sug\-gests that one alternatively considers the grand-canonical ensemble,
where the fugacity~$z$ is fixed,
whereas the total number~$N$ of particles in the system fluctuates.
In that ensemble the stationary distribution of the occupations
is strictly a product measure:
\beq
P\GC(\{N_m\})=\frac{1}{\Z_M(z)}\prod_{m=1}^M (zq_m)^{N_m}\,p_{N_m}
=\prod_{m=1}^M f\GC_{m,k},
\label{pgcfac}
\eeq
where the normalized factor $f\GC_{m,k}$
gives the distribution of the occupation of site~$m$:
\beq
f\GC_{m,k}=\prob{N_m\GC=k}=\frac{(zq_m)^k\,p_k}{P(zq_m)}.
\label{fgc}
\eeq
Thus the mean occupation at site $m$ reads
\beq
\rho\GC_m=\mean{N_m\GC}=\sum_{k\ge0} kf\GC_{m,k}=\frac{zq_mP'(zq_m)}{P(zq_m)}
=z\frac{\partial}{\partial z}\ln P(zq_m),
\label{rhogc}
\eeq
where the accent denotes differentiation, and finally the mean density of the system is
\beq
\rho\GC=\frac{\mean{N\GC}}{M}=\frac{1}{M}\sum_{m=1}^M\rho\GC_m
=\frac{z\Z'_M(z)}{M\Z_M(z)}=\frac{z}{M}\frac{\partial}{\partial z}\ln\Z_M(z).
\label{mrhogc}
\eeq

\subsection{The case of independent random walkers ($u_k=k$).}

We now turn to the simplest example of a ZRP, corresponding to $u_k=k$.
As recalled in the introduction,
this situation describes independent random walkers.
We have
\beq
p_k=\frac{1}{k!},\qquad P(z)=\e^z.
\eeq

\subsubsection*{Grand-canonical ensemble.}

The grand partition function reads
\beq
\Z_M(z)=\e^{Mz\overline{q}},
\eeq
where we have introduced the notation
\beq
\overline{q}=\frac{1}{M}\sum_{m=1}^M q_m
\label{qbardef}
\eeq
for the mean single-particle weight.
The occupations therefore follow the Poissonian law
\beq
f\GC_{m,k}=\e^{-zq_m}\frac{({zq_m})^k}{k!},
\eeq
and the local densities read
\beq
\rho_m\GC=zq_m.
\eeq
Thus the mean density and the fugacity are related through
\beq
\rho\GC=z\overline{q}.
\eeq
As a consequence, for any density $\rho$, there exists a value of the fugacity,
namely $z=\rho/\overline{q}$, such that $\rho\GC(z)=\rho$.
The system can sustain any density of particles and is therefore always in a fluid phase.

\subsubsection*{Canonical ensemble.}

The canonical partition function is
\beq
Z_{M,N}=\frac{(M\overline{q})^N}{N!},
\label{zclassic}
\eeq
and the local densities read
\beq
\rho_m=\frac{N q_m}{M\overline{q}}.
\label{rhoclassic}
\eeq
The canonical joint distribution of the occupations is multinomial:
\beq
P(\{N_m\})=\frac{1}{(M\overline{q})^N}
\;\frad{N!}{\prod_{m=1}^M N_m!}
\;\prod_{m=1}^M q_m^{N_m},
\label{multinomial}
\eeq
as it should be,
since independent random walkers behave as independent classical particles.

\subsection{The occupation-independent case ($u_k=1$).}

The occupation-independent ZRP with $u_k=1$ is another simple example, yet
richer because the random walkers are no longer independent.
We have then
\beq
p_k=1,\qquad P(z)=\frac{1}{1-z}.
\eeq

\subsubsection*{Grand-canonical ensemble.}

The grand partition function reads
\beq
\Z_M(z)=\prod_{m=1}^M\frac{1}{1-z q_m}.
\eeq
The occupations therefore follow the geometric law
\beq
f\GC_{m,k}=(1-zq_m)(zq_m)^k,
\label{ffree}
\eeq
and the local densities read
\beq
\rho\GC_m=\frac{zq_m}{1-zq_m}.
\label{rhofree}
\eeq
Thus the mean density and the fugacity are related through
\beq
\rho\GC=\frac{1}{M}\sum_{m=1}^M\frac{zq_m}{1-zq_m}.
\label{mrhofree}
\eeq

For a homogeneous system ($q_m=1$), this relation simplifies to
\beq
\rho\GC=\frac{z}{1-z},
\eeq
which has a solution $z(\rho\GC)$ for any value of the density, and hence the
system is again in a fluid phase.
However, if the system is inhomogeneous,
local densities are given by~(\ref{rhofree}), and a condensation transition may occur.
The simplest case is that of an attractive impurity~\cite{braz,angel},
where one of the $q_m$ is larger than the common value of all the other ones.
A condensation transition also occurs in the thermodynamic limit
when the $q_m$ are random and distributed according to~(\ref{fdef}) with $c>1$.
The latter phenomenon is addressed in section~\ref{thfree}.

It is worth emphasizing the analogy between the statics of the occupation-independent ZRP
and that of an ideal Bose gas~\cite{hanney,evans96}.
Setting $z=\e^{\beta\mu}$,
where~$\beta$ is the inverse temperature and $\mu$ the chemical potential,
and $q_m=\e^{-\beta E_m}$,
the expression~(\ref{rhofree}) for the local particle density at site $m$
identifies with the Bose occupation factor
\beq
f_{\rm Bose}(E_m)=\frac{1}{\e^{\beta(E_m-\mu)}-1}
\label{fbose}
\eeq
of a fictitious quantum-mechanical level with energy $E_m$.
Identifying the connection between classical stochastic processes and the ideal Bose gas
dates back to~\cite{leeuw}.

The distribution~(\ref{fdef}) of the single-particle weights
translates to a distribution of energy levels (density of states) $D(E)$
vanishing as $D(E)\sim E^{c-1}$ as the energy $E>0$ approaches zero.
The latter power-law singularity is to be put in perspective
with the density of states of a free particle in a box in $d$-dimensional space,
i.e., $D(E)\sim E^{d/2-1}$.
This parallel leads to the identification $c=d/2$ between the disorder exponent $c$
and half the dimensionality of the space where the Bose gas lives.
The condition $c>1$ for having a condensation transition~\cite{krug96,evans96},
recalled above and to be worked out in section~\ref{thfree},
is therefore equivalent to the well-known result
that an ideal Bose gas exhibits
a Bose-Einstein condensation in dimension $d>2$ only.

\subsubsection*{Canonical ensemble.}

The canonical partition function can be evaluated
from the representation~(\ref{contour}) as
\beq
Z_{M,N}=\oint\frac{\d z}{2\pi\ii z^{N+1}}
\prod_{m=1}^M\frac{1}{1-zq_m}.
\label{cfree}
\eeq
Evaluating the contour integral by the method of residues, we get
\beq
Z_{M,N}=\sum_{m=1}^Mq_m^NQ_m,
\label{zcfree}
\eeq
with
\beq
Q_m=\prod_{n\ne m}\frac{q_m}{q_m-q_n}.
\eeq
The local densities can then be obtained
by differentiating the above formulas according to~(\ref{rhores}).
We thus obtain
\beq
\rho_m=\frac{1}{Z_{M,N}}
\left(Nq_m^NQ_m+\sum_{n\ne m}\frac{q_nq_m^NQ_m+q_mq_n^NQ_n}{q_n-q_m}\right).
\label{rcfree}
\eeq

The expressions~(\ref{zcfree}) and~(\ref{rcfree}) are much more intricate
than their counter\-parts~(\ref{zclassic}) and~(\ref{rhoclassic})
in the case of independent random walkers.
This difference can be attributed to the fact that the canonical ensemble
is the natural framework to describe identical classical particles
(hence the simplicity of the multinomial joint distribution~(\ref{multinomial})),
whereas the grand-canonical ensemble is the only natural one when dealing with
quantum-mechanical energy levels
(hence the analogy between the expression~(\ref{rhofree})
for the grand-canonical local density
and the Bose occupation factor~(\ref{fbose})).

\subsection{On the choice of single-particle weights made in this work}
\label{ourmodel}

Let us come back to the specific model considered in the present work.
For a given ZRP, defined by the rates $u_k$ and $w_{mn}$,
the stationary properties only rely on the knowledge of the weights $p_k$ and $q_m$.
The former ones are easily deduced from the rates $u_k$~(see~(\ref{pk})).
In contrast, the stationary master equation~(\ref{master}) cannot be solved in
closed form in general.
The problem indeed belongs to the rich and complex area
of diffusion in random media~\cite{drm}.
In this work we have adopted the alternative viewpoint of modeling diffusion disorder
by considering the single-particle weights $q_m$ as i.i.d.~random variables
drawn from the distribution~(\ref{fdef}).

Let us take the example of the mean-field geometry of a completely connected
graph over $M$ sites to discuss the issue.
Consider the inhomogeneous diffusion where
the particle leaves site $m$ with some site-dependent rate $\lambda_m$,
and jumps to a random arrival site~$n$, chosen uniformly over the system.
We have therefore $w_{mn}=\lambda_m/M$.
It can be checked that the stationary probabilities read $q_m=1/\lambda_m$,
up to normalization.
In the more general situation
of separable rates of the form $w_{mn}=\lambda_m\mu_n$,
we have $q_m=\mu_m/\lambda_m$, again up to normalization.
The stationary state thus obtained is an equilibrium state,
obeying the condition of detailed balance.
The mean current from site $m$ to site $n$, $J_{mn}=w_{mn}q_m=\mu_m\mu_n$,
up to normalization, is indeed symmetric.
Another example is the case of the totally asymmetric diffusion on a
one-dimensional lattice, either infinite or finite with periodic boundary conditions,
such that $w_{mn}=\lambda_m\,\delta_{n,m+1}$.
Here again, we have $q_m=1/\lambda_m$, up to normalization.

In the two above examples of inhomogeneous diffusion,
the single-particle weights~$q_m$ can be viewed as independent random variables,
and they can take any prescribed values.
This somehow justifies our choice of modeling them by i.i.d.~random variables
with an arbitrary distribution $f(q)$.
It should however be clear that the single-particle weights
cannot be viewed as independent random variables in the most general situation.
Consider for definiteness a finite-dimensional lattice with weakly disordered
asymmetric hopping rates $w_{mn}$ on its bonds.
In this case, the $q_m$ are given by the normalized solution of
the stationary master equation~(\ref{master}),
which may exhibit a non-trivial spatial structure,
especially in the one-dimensional case of the Sinai model~\cite{drm}.

Finally, the rationale for the choice~(\ref{fdef}) of the distribution $f(q)$
of single-particle weights is as follows.
The properties of the condensate are dictated by the largest single-particle weights,
and therefore by the behavior of the distribution $f(q)$ near its maximum $q_\max$.
The situation where $q_\max$ is infinite is somehow pathological
and requires a case-by-case study.
Section~\ref{networks} contains an investigation of a physically motivated situation
of this type, namely complex networks.
In the case where $q_\max$ is finite (and can then be set equal to unity),
the situation of a power-law singularity turns out to be the appropriate framework
to disclose the phase diagram.
If the distribution $f(q)$ vanishes faster than a power law as $q\to q_\max=1$,
then one has formally $c=\infty$,
thus, for $\rho>\rho_c$ the system will always be in the localized condensed phase,
irrespective of the interaction exponent $b$.
Finally, the pure power law~(\ref{fdef}) has been chosen merely for simplicity.

\section{Two limiting case studies}
\label{limits}

As mentioned in the introduction,
the full model defined by~(\ref{fdef}) and~(\ref{rate}) contains
two limiting prototypical cases
which are known to exhibit a condensation transition.
We recall here the properties of these models
which will be relevant for the study of the full model,
to be presented in section~\ref{thermo}.

\subsection{Homogeneous ZRP: spontaneous symmetry breaking}
\label{homo}

The homogeneous ZRP
(see~\cite{braz,hanney,lux} for reviews)
corresponds to the case where the single-particle weights~$q_m$ are uniform,
i.e., do not depend on the site.
This homogeneity property holds for the complete graph with uniform hopping rates,
and more generally whenever all sites are symmetry-related and thus equivalent.
This is automatically satisfied if the sites are related to each other
by translation invariance,
like e.g.~for usual symmetric or biased random walk
on finite lattices with periodic boundary conditions.

Consider the homogeneous ZRP with rate~(\ref{rate}).
Setting $q_m=1$, we have
\beq
\Z_M(z)=P(z)^M.
\eeq
As a consequence, in the grand-canonical ensemble,
(\ref{fgc}) and~(\ref{rhogc}) respectively become
\beq
f\GC_k=\frac{z^kp_k}{P(z)}
\label{fhomo}
\eeq
and
\beq
\rho\GC=\frac{zP'(z)}{P(z)}.
\label{rhomo}
\eeq

In the canonical ensemble, in the thermodynamic limit
($M\to\infty$, $N\to\infty$, $\rho=N/M$ fixed),
the condition~(\ref{rhomo}) is recovered
by estimating the contour integral~(\ref{contour}) by the saddle-point method.
This shows that both statistical ensembles are equivalent in the thermodynamic limit
as long as the system remains fluid, i.e., provided no condensate appears,
as could be expected.

With the rate~(\ref{rate}), we have
\beq
p_k=\frac{\Gamma(b+1)\,k!}{\Gamma(k+b+1)}
=\int_0^1 u^k\,b(1-u)^{b-1}\,\d u\approx\frac{\Gamma(b+1)}{k^b},
\label{pkas}
\eeq
hence
\beq
P(z)=\int_0^1\frac{b(1-u)^{b-1}}{1-zu}\,\d u=\null_2F_1(1,1;b+1;z),
\eeq
where $_2F_1$ is the hypergeometric function.

The following values will be needed in the sequel:
\beqa
&&P(1)=\frad{b}{b-1},\qquad
P'(1)=\frad{b}{(b-1)(b-2)},\nonumber\\
&&P''(1)=\frad{4b}{(b-1)(b-2)(b-3)},
\label{part}
\eeqa
where it is understood that these quantities are convergent
respectively for $b>1$, $b>2$, and $b>3$.
More generally, the function $P(z)$ has a branch cut at $z=1$,
with a singular part of the form
\beq
P_\sg(z)\approx\frad{\pi b}{\sin\pi b}\,(1-z)^{b-1}.
\label{psg}
\eeq
Whenever $b$ is a positive integer or a half-integer,
$P(z)$ can be expressed in terms of elementary functions.
In particular, for every integer $b=n\ge1$,
its singular part is of the form
$P_\sg(z)\approx(-1)^nn(1-z)^{n-1}\ln(1-z)$.
We have, for instance,
\beqa
b=1&:&\quad P(z)=-\frac{\ln(1-z)}{z},\nonumber\\
b=2&:&\quad P(z)=\frac{2(1-z)\ln(1-z)}{z^2}+\frac{2}{z},\nonumber\\
b=3&:&\quad
P(z)=-\frac{3(1-z)^2\ln(1-z)}{z^3}-\frac{3}{z^2}+\frac{9}{2z},\nonumber\\
b=4&:&\quad
P(z)=\frac{4(1-z)^3\ln(1-z)}{z^4}+\frac{4}{z^3}-\frac{10}{z^2}+\frac{22}{3z}.
\label{pex}
\eeqa

The homogeneous ZRP with rate~(\ref{rate}) has a continuous phase transition
when the fugacity $z$ reaches the singular point $z_c=1$.
For $b>2$, this takes place at a finite critical density (see~(\ref{rhoc1}))
\beq
\rho_c=\frac{P'(1)}{P(1)}=\frac{1}{b-2}.
\label{rhoch}
\eeq
This critical density separates a fluid phase $(\rho<\rho_c)$
and a condensed phase $(\rho>\rho_c)$,
whose main characteristics are as follows.

\begin{itemize}

\item
{\it Fluid phase $(\rho<\rho_c)$.}
The canonical and grand-canonical ensembles are equivalent throughout the fluid phase.
The density $\rho=\rho\GC$
increases from 0 to $\rho_c$ as the fugacity $z$ increases from 0 to 1.
The occupation probabilities $f_k=f_k\GC$ (see~(\ref{fhomo})) fall off exponentially.

\item
{\it Critical density $(\rho=\rho_c)$.}
This density is reached when the fugacity $z$ takes the singular value $z_c=1$.
It is therefore the maximal density that can be reached
in the grand-canonical ensemble.

At the critical density, the occupation probabilities
\beq
f_k=\frac{p_k}{P(1)}\approx\frac{(b-1)\Gamma(b)}{k^b}
\label{afkc}
\eeq
fall off as a power law with exponent $b$.
The second moment of the occupation probabilities at criticality,
\beq
\mu_c=\sum_{k\ge0}k^2f_k=\frac{P'(1)+P''(1)}{P(1)}
=\frac{b+1}{(b-2)(b-3)},
\eeq
is convergent for $b>3$ (regime of normal fluctuations),
whereas it is divergent for $2<b<3$ (regime of anomalous fluctuations).

\item
{\it Condensed phase $(\rho>\rho_c)$.}
The condensed phase only exists in the canonical ensemble
where the total number $N$ of particles is imposed.
A large ZRP in its condensed phase
consists of a uniform critical background,
characterized by the critical occupation probabilities~(\ref{afkc}),
and of a single macroscopic condensate,
containing an extensive number of excess particles of order $\Delta=M(\rho-\rho_c)$
(see~(\ref{ddef})).
For a system where all sites are equivalent,
the condensate can be at any site $m$ with probability $1/M$.
The fluctuations of the number of particles in the condensate
around $\Delta$ are known to be Gaussian and to scale as $M^{1/2}$
in the regime of normal fluctuations ($b>3$),
whereas they have a broad distribution and scale as $M^{1/(b-1)}$
in the regime of anomalous fluctuations ($2<b<3$)~\cite{gross,maj}.

\end{itemize}

For the homogeneous ZRP,
the spontaneous symmetry breaking (SSB) scenario in the condensed phase,
with its unique condensate on top of a critical background in the thermodynamic limit,
has been proved rigorously~\cite{gross,armendariz}.
The rare configurations where the condensate is not unique,
i.e., where the $\Delta$ excess particles are shared by two sites
in significant proportions,
have been argued to occur with a probability falling off as
the power law $1/M^{b-2}$~\cite{us}.

Let us close with a comment on universality.
The condition $b>2$ for the existence of a condensed phase is universal,
in the sense that any rate with asymptotic behavior $u_k=1+b/k+\cdots$
yields a finite critical density~$\rho_c$ only for $b>2$.
In contrast, the value of $\rho_c$ is not universal
as it depends on the full shape of the rate $u_k$.
Let us mention e.g.~that the~rate~\cite{lux,cg}
\beq
u_k=\left(1+\frac{1}{k}\right)^b
\label{zetarate}
\eeq
yields $p_k=(k+1)^{-b}$ and $\rho_c=\zeta(b-1)/\zeta(b)$,
where $\zeta$ is Riemann's zeta function.
This choice of rate makes sense for all values of $b$,
whereas~(\ref{rate}) is limited to the range $b>-1$ (as $u_1=b+1>0$).
The stationary state of the ZRP with this rate coincides
with the static model analyzed in~\cite{bia}.

\subsection{Occupation-independent inhomogeneous ZRP: explicit symmetry breaking}
\label{thfree}

We now consider the occupation-independent inhomogeneous ZRP ($u_k=1$),
where the single-particle weights $q_m$ are i.i.d.~random variables
drawn from the distribution~(\ref{fdef}).

The analysis of the model~\cite{krug96,evans96}
is again simpler in the grand-canonical ensemble.
Taking the thermodynamic limit of~(\ref{mrhofree}), we readily obtain
\beq
\overline\rho\GC=\int_0^1\frac{zq}{1-zq}\,f(q)\,\d q.
\eeq
The critical density $\rho_c$ is then obtained
by evaluating the above expression at the critical fugacity $z_c=1/q_\max=1$:
\beq
\rho_c=\int_0^1\frac{q}{1-q}\,f(q)\,\d q.
\eeq
This integral is convergent for $c>1$.
It should however be underlined that the existence of a finite
critical density $\rho_c$ is a characteristic of the thermodynamic limit.
The density~(\ref{mrhofree}) of any finite system indeed diverges,
albeit at a further point, i.e., $z\to1/q_1$, where $q_1<1$
denotes the largest single-particle weight.
For the distribution~(\ref{fdef}) we thus get (see~(\ref{rhoc2}))
\beq
\rho_c=\frac{1}{c-1}.
\eeq

The mean occupation distribution at the critical density
is obtained by averaging the geometric law~(\ref{ffree})
over the distribution $f(q)$, again at $z=z_c=1$.
It reads
\beq
\bar f_k=\int_0^1(1-q)q^k\,f(q)\,\d q
=\frac{c\Gamma(c+1)k!}{\Gamma(k+c+2)}\approx\frac{c\Gamma(c+1)}{k^{c+1}}.
\eeq
We thus recover the property that the critical density
\beq
\rho_c=\sum_{k\ge0}k\,\bar f_k
\eeq
is convergent for $c>1$.
Finally, the mean square critical occupation
\beq
\mu_c=\sum_{k\ge0}k^2\bar f_k
=\int_0^1\frac{q(1+q)}{(1-q)^2}\,f(q)\,\d q=\frac{c+2}{(c-1)(c-2)}
\eeq
is convergent for $c>2$.

In the canonical ensemble, there is again a condensed phase for $\rho>\rho_c$.
In the latter phase, a large but finite system typically has
a single condensate of size $\Delta$ (see~(\ref{ddef})).
This condensate sits on the site with the largest single-particle weight
($q_1$, say) with very high probability.
This almost sure localization of the condensate on the most favored site
is the gist of the explicit symmetry breaking (ESB) scenario
at work in the occupation-independent inhomogeneous ZRP
(see section~\ref{thloc} for more details).

The exponent $c$ characterizing diffusion disorder therefore plays, in the
present situation of ESB, a role
analogous to that of the exponent $b$ characterizing interactions
in the situation of SSB on a homogeneous system.
We have a fluid phase for $c<1$ or $b<2$,
a condensed phase with anomalous fluctuations for $1<c<2$ or $2<b<3$,
and a condensed phase with normal fluctuations for $c>2$ or $b>3$.

\section{Universal phase diagram}
\label{thermo}

We now address the study of the stationary state of the full inhomogeneous ZRP,
characterized by the interaction exponent $b$ entering the rate~(\ref{rate})
and by the disorder exponent $c$ entering the distribution~(\ref{fdef})
of the single-particle weights.
In this section, we aim at constructing the universal phase diagram of this ZRP
in the plane of the exponents $b$ and $c$,
displayed in figures~\ref{phase1} and~\ref{phase2}.

\subsection{Existence and nature of the condensate
(critical density, critical fluctuations)}
\label{thgal}

The analysis is again easier within the grand-canonical approach.
The mean occupation profile $\bar f_k$ at the critical density is obtained by
averaging~(\ref{fgc}) over the distribution $f(q)$
at the critical fugacity $z_c=1$.
We thus obtain
\beq
\bar f_k=p_k\int_0^1\frac{q^k}{P(q)}\,f(q)\,\d q,
\label{gbarf}
\eeq
and hence
\beqa
\rho_c&=&\int_0^1\frac{qP'(q)}{P(q)}\,f(q)\,\d q,\nonumber\\
\mu_c&=&\int_0^1\frac{qP'(q)+q^2P''(q)}{P(q)}\,f(q)\,\d q.
\label{grhomuc}
\eeqa

If the critical density $\rho_c$ is finite,
the model has a condensed phase.
On a large but finite system of $M$ sites,
if the particle density $\rho=N/M$ exceeds $\rho_c$,
an extensive number $\Delta=M(\rho-\rho_c)$ of excess particles
(see~(\ref{ddef})) will typically form a unique condensate.
Furthermore the finiteness of $\mu_c$ will determine whether the size fluctuations
of the condensate around $\Delta$ are normal or not.

The uniqueness of the condensate,
which has been proved rigorously in the homogeneous case~\cite{gross,armendariz},
is basically due to the fact that
the convexity of the interactions between particles
suppresses the mixed configurations where the condensate is not unique,
i.e., where the $\Delta$ excess particles are shared by two or more sites
in significant proportions.
This phenomenon is so general that it can already be exemplified
on a ZRP on two sites for any $b>0$
(see figure~\ref{occtwosites} in~\ref{sectwosites}).
Let us mention that a rigorous combinatorial analysis
of the condensation phenomenon in a finite system
has been reported in~\cite{fls}.

It is therefore of central importance to control the convergence properties
of the integral expressions~(\ref{grhomuc}) for $\rho_c$ and $\mu_c$.
These properties are entirely dictated by the regime $q\to1$.
Two cases have to be dealt with separately.

\begin{itemize}

\item
For $b>1$, $P(1)=b/(b-1)$ is finite~(see~(\ref{part})).
As a consequence, the integral entering~(\ref{gbarf}) falls off as $k^{-c}$.
This observation yields the power-law decay
\beq
\bar f_k\sim\frac{1}{k^{b+c}}\qquad(b>1).
\label{eq:fk1}
\eeq
Thus $\rho_c$ is finite for $b+c>2$, while $\mu_c$ is finite for $b+c>3$.
These two properties can be directly checked by analyzing the integrals~(\ref{grhomuc}).
They generalize the well-known conditions $b>2$ and $b>3$ for the homogeneous case,
which are consistently recovered in the $c\to0$ limit.

\item
For $b<1$, $P(q)\sim(1-q)^{b-1}$ is divergent~(see~(\ref{psg})).
As a consequence, the integral entering~(\ref{gbarf}) falls off as $k^{b-c-1}$.
We thus obtain the power-law decay
\beq
\bar f_k\sim\frac{1}{k^{c+1}}\qquad(b<1).
\label{eq:fk2}
\eeq
As a consequence, $\rho_c$ is finite for $c>1$, while $\mu_c$ is finite for $c>2$.
These conditions, which were already derived in section~\ref{limits}
in the occupation-independent case ($u_k=1$),
thus extend to the whole range $b<1$.
They can again be directly checked by analyzing the integrals~(\ref{grhomuc}).

\end{itemize}

To sum up,
the inhomogeneous ZRP defined by~(\ref{fdef}) and~(\ref{rate}) exhibits a condensation transition
only if attractive interactions and/or diffusion disorder
excess a threshold value.
The critical density $\rho_c$ is indeed finite
for all $c>0$ if $b>2$, for $c>2-b$ if $1<b<2$, and for $c>1$ if $b<1$.
Similarly, the critical mean square occupation $\mu_c$ is finite
for all $c>0$ if $b>3$, for $c>3-b$ if $1<b<3$, and for $c>2$ if $b<1$.
The above inequalities demarcate the different regions
of the phase diagrams of figures~\ref{phase1} and~\ref{phase2}.

Whenever the critical density $\rho_c$ is finite,
it is a rapidly decreasing function of the exponents~$b$ and $c$.
The estimate
\beq
\rho_c\approx\frac{1}{bc}
\eeq
holds whenever $b$ and $c$ are simultaneously large.
In the regime where $c$ is small, for any $b>2$,
the critical density departs linearly from its value $\rho_c(0)=1/(b-2)$
in the homogeneous case (see~(\ref{rhoc1}),~(\ref{rhoch})), as
\beq
\rho_c(c)=\rho_c(0)(1-\I c+\cdots),
\eeq
with
\beq
\I=\int_0^1\left(1-\frac{qP'(q)}{\rho_c(0)\,P(q)}\right)\frac{\d q}{1-q}.
\label{inti}
\eeq

\begin{figure}[!ht]
\begin{center}
\includegraphics[angle=-90,width=.5\linewidth]{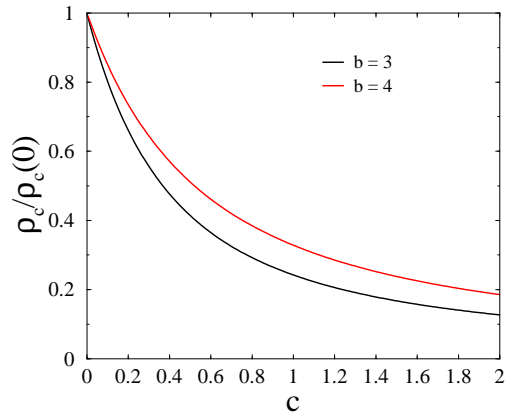}
\caption{\small
Critical density $\rho_c$,
normalized by $\rho_c(0)$, against the diffusion disorder exponent $c$,
for $b=3$ and $b=4$.}
\label{rhoc}
\end{center}
\end{figure}

Figure~\ref{rhoc} shows a plot of the ratio $\rho_c/\rho_c(0)$ against $c$
for $b=3$ and $b=4$.
For these integer values of the exponent $b$,
the function $P(z)$ admits the explicit forms~(\ref{pex}),
which ease the numerical evaluation of the integrals~(\ref{grhomuc}) and~(\ref{inti}).
For $b=3$, we have $\rho_c(0)=1$ and $\I=2.34704$.
For $b=4$, we have $\rho_c(0)=1/2$ and $\I=1.70735$.

\subsection{Localization properties of the condensate}
\label{thloc}

It is intuitively clear that the site with the largest single-particle weight
($q_1$, say) has the highest chance of hosting the condensate.
However, if the spacings between the largest weights $q_m$ are small enough, it is
plausible to expect a competition amongst the corresponding sites to host the condensate.
The aim of this section is to provide a quantitative treatment of this issue.

We first need to characterize the law of spacings between the $q_m$.
This can be achieved by using the order statistics of the original unordered
variables.
Let us reorder the sites according to decreasing single-particle weights $q_m$
and relabel them so as to have $q_1>q_2>\dots>q_M$.
A first estimation of the $m$-th ordered weight $q_m$ can be obtained by equating
the ratio $m/M$
and the cumulative probability for the random variable $q$ to be larger than $q_m$,
i.e., $F(q_m)$, with
\beq
F(q)=\int_q^1f(q')\,\d q'=(1-q)^c.
\label{fcumul}
\eeq
We thus obtain the estimate
\beq
1-q_m\approx\left(\frac{m}{M}\right)^{1/c}.
\label{rough}
\eeq
A more precise prediction can be obtained as follows.
In the regime of most interest where $M$ is large, while the order $m$ is kept finite,
the set of rescaled variables
\beq
x_m=M F(q_m)=M(1-q_m)^c
\label{xdef}
\eeq
is asymptotically distributed as Poissonian points
with unit density on the positive real line.
This property is clear for $c=1$, where the $q_m$ are uniformly distributed,
so that $F(q)=1-q$.
In this case, the variables $x_m$ are obtained by ordering
$M$ points drawn independently and uniformly in the large interval $[0,M]$.
The sequence thus obtained converges to Poissonian points with unit density.
The general case then follows by changing variables from $q$ to $1-F(q)$.
A full account of the connections
between extreme-value statistics or order statistics and point processes
can be found in~\cite[Sec.~7.3]{orderstats}.

For Poissonian points with unit density, we have
\beq
x_m=\tau_1+\tau_2+\cdots+\tau_m,
\label{xrenewal}
\eeq
where the distances $\tau_m=x_m-x_{m-1}$ between successive points
are i.i.d.~exponential variables with the distribution
$f_\tau(\tau)=\e^{-\tau}$ (and $x_0=0$).
We thus obtain the more precise estimate
\beq
1-q_m\approx\left(\frac{x_m}{M}\right)^{1/c},
\label{order}
\eeq
where the distribution of $x_m$ is a Gamma distribution $\Gamma(m,1)$:
\beq
f_{x_m}(x)=\frac{x^{m-1}\,\e^{-x}}{(m-1)!}.
\eeq
The mean and the variance of $x_m$ thus read
\beq
\overline{x_m}=\var x_m=m.
\eeq
So, as the order $m$ gets large,
the distribution of $x_m$ becomes peaked around $m$,
with Gaussian fluctuations of relative order $1/\sqrt{m}$.
In this regime~(\ref{order}) simplifies to~(\ref{rough}).

We can now turn to the localization properties
of the condensate in the stationary state of a large but finite system.
This aspect of the problem is more subtle than the issue discussed in section~\ref{thgal}.
Indeed, a condensate can only be dealt with in the canonical ensemble at fixed $N$,
which is less amenable to analytical studies.
Moreover, we are interested in properties of single sites ($m=1$, 2, $\dots$)
and not in averaged properties over the whole system.
We therefore have to rely on the following more heuristic line of reasoning.

Consider a large but finite system in its condensed phase ($M\gg1$, $\rho>\rho_c$),
for $b>0$ and for a given draw of the single-particle weights $q_m$.
We already know the following:
in a typical configuration, the condensate is unique,
and its size fluctuations around $\Delta=M(\rho-\rho_c)$ are negligible.
Our goal is to estimate the probability~$\Pi_m$ that site $m$ hosts the condensate
in the stationary state.
The expression~(\ref{pfac}) of the weight of an arbitrary configuration
shows that this probability (unnormalized so far) can be estimated by
\beq
\Pi_m\sim q_m^\Delta,
\label{pieps}
\eeq
hence, using~(\ref{order}),
\beq
\Pi_m\sim\exp\left(-\Delta\left(\frac{x_m}{M}\right)^{1/c}\right).
\label{pisca}
\eeq
The argument of the exponential is proportional to $\Delta/M^{1/c}\sim M^{1-1/c}$.
This estimate singles out the borderline value $c=1$,
where the dependence on the system size $M$ drops out.
This phenomenon already takes place in the simple case of the ZRP on two sites.
In this situation, analyzed in~\ref{sectwosites},
quantities of interest
depend on the rescaled inhomogeneity parameter $\theta=N\eps$ (see~(\ref{thdef})).
In the present case of a large system, the role of $\eps$ is played by
the difference between the largest two weights,
i.e., $q_1-q_2\sim M^{-1/c}$ (see~(\ref{order})),
so that $\theta\sim\Delta/M^{1/c}\sim M^{1-1/c}$
scales exactly as the argument of the exponential in~(\ref{pisca}).

The following regimes have to be dealt with separately,
according to the value of the exponent $c$ with respect to the borderline value $c=1$.

\begin{itemize}

\item
{\it Extended regime} ($c<1$).
In this first regime,
$\rho_c$ is finite for $b>2-c$.
The distribution $f(q)$ of the single-particle weights
diverges as $q\to q_\max=1$.
Hence the few largest weights $q_1$, $q_2$, $q_3,\dots$
pile up near $q=1$.
The typical distance between them, scaling as $M^{-1/c}$,
is much smaller than the mean spacing $1/M$ in the bulk of the distribution.
As a consequence, the probability $\Pi_m$
for site~$m$ to host the condensate
takes appreciable values for many values of~$m$.
In the regime of interest, i.e., $1\ll m\ll M$, we have $x_m\approx m$,
and so the estimate~(\ref{pisca}) simplifies~to
\beqa
\Pi_m
&\sim&\exp\left(-\Delta\left(\frac{m}{M}\right)^{1/c}\right)
\nonumber\\
&\sim&\exp\left(-(\rho-\rho_c)M^{1-1/c}m^{1/c}\right).
\label{piasy}
\eeqa
The typical number $R$ of favored sites can now be estimated as
the range of values of $m$ for which $\Pi_m$ remains comparable to $\Pi_1$.
We are thus led to estimate $R$ as
the value of~$m$ such that the {\it argument} of the exponential function
in~(\ref{piasy}) is of order unity.
We thus obtain
\beq
R\sim\frac{M}{\Delta^c}=\frac{M^{1-c}}{(\rho-\rho_c)^c}.
\label{kres}
\eeq
In a typical configuration of the stationary state of the system,
there is a unique condensate sitting at a well-defined site.
The probabilities $\Pi_m$ that the various sites $m$ host the condensate
are however extended over a large {\it hosting set} of favored sites.
The typical size $R$ of this set
grows according to the sub-extensive law~(\ref{kres}).
In the $c\to0$ limit we recover an extensive growth $R\sim M$,
in agreement with the fully extended nature of the condensate
in the homogeneous ZRP,
with its spontaneous symmetry breaking mechanism.
In the opposite regime ($c\to1$), the result~(\ref{kres}) crosses over
to a localized condensate, in agreement with an explicit symmetry breaking mechanism.
To sum up, the extended condensed phase,
with its sub-extensive hosting set,
realizes a continuous interpolation
between the two condensation scenarios which were known so far.

\item
{\it Borderline case} ($c=1$).
In the borderline situation corresponding to the critical value $c=1$
(and so $\rho_c$ is finite for $b>1$),
any dependence on the system size drops out of the estimate~(\ref{pisca}),
which reads
\beq
\Pi_m\sim\e^{-(\rho-\rho_c)x_m},
\label{pic}
\eeq
where the $x_m$ are Poissonian points with unit density.
The probabilities $\Pi_m$ therefore typically take appreciable values
on a small and fluctuating number of sites.
\ref{borderline} is devoted to a detailed study of these fluctuating probabilities.
Distributions of this sort seem to have been first described
in the context of randomly breaking an interval~\cite{df}.
They have since then been met in several circumstances.
A useful tool to investigate them is provided by the quantity $Y$ (see~(\ref{ydef})),
which is somehow similar to the participation ratio
used in the theory of Anderson localization~\cite{ipr}.
In the present situation,
this quantity $Y$ will be shown to have a non-trivial distribution,
which is universal in the sense that it only depends on the parameter $r=\rho-\rho_c$.

\item
{\it Localized regime} ($c>1$).
For $c>1$ (and so $\rho_c$ is now finite for all $b$),
the distribution $f(q)$ goes to zero as $q\to q_\max=1$.
The few largest weights $q_1$, $q_2$, $q_3,\dots$
are therefore rather distant from one another.
The successive distances between them indeed scale as $M^{-1/c}$.
As a consequence, for a typical draw of the single-particle weights,
the first probability $\Pi_1$ is overwhelmingly larger than the other ones.
The condensate is therefore localized on the most favored site ($m=1$)
with very high probability.
The events where the largest two weights
$\Pi_1$ and $\Pi_2$ become comparable
occur with a small probability,
which can be estimated to be of order $M^{-(1-1/c)}$.
These rare events will be analysed more precisely in~\ref{effloc}.

\end{itemize}

\subsection{Phase diagram}

Putting all the above results together,
we arrive at the universal phase diagram depicted in figures~\ref{phase1}
and~\ref{phase2}.
The phase diagram shown in figure~\ref{phase1}
exhibits three phases
in the plane of the exponents~$b$ (characterizing interactions)
and~$c$ (characterizing diffusion disorder).
In the fluid phase, there is no condensate at any density.
In the localized condensed phase,
for $\rho>\rho_c$, the condensate lives on the most favored site ($m=1$)
with very high probability.
In the extended condensed phase,
for $\rho>\rho_c$, the condensate lives on one of the sites of a large set of
favored sites, whose size~$R$ grows sub-extensively, according to~(\ref{kres}).
Figure~\ref{phase2} shows a more detailed phase diagram,
where the condensed phases are further separated into
phases where $\mu_c$ is finite,
hence the size fluctuations of the condensate around $\Delta$
are normal (i.e., Gaussian and growing as $M^{1/2}$),
and phases where~$\mu_c$ is divergent,
hence those fluctuations are anomalous, i.e., not Gaussian and growing faster
than $M^{1/2}$.
More precisely, these fluctuations scale as $M^{1/c}$ for $b<1$ and $1<c<2$,
and as $M^{1/(b+c-1)}$ for $b>1$ and $2-b<c<3-b$.
These growth laws can be respectively deduced
from the expressions~(\ref{eq:fk2}) and~(\ref{eq:fk1}).

It should be clear from its derivation
that the above phase diagram is universal, as announced in the introduction,
in the sense that it holds
for any rate whose asymptotic behavior at large $k$
is of the form $u_k\approx1+b/k$,
hence $P_\sg(z)\sim(1-z)^{b-1}$,
and for any distribution of single-particle weights with a finite maximum $q_\max$
and a power-law scaling as $f(q)\sim(q_\max-q)^{c-1}$.

\section{Numerical illustrations}
\label{numerics}

This section is devoted to numerical illustrations of the predictions made in the previous section.

\subsection{Density and occupation probability profiles}

Let us start at a rather qualitative level,
and look at the growth of the density profile across a finite system
as particles are added one by one.
In order to avoid irregularities due to fluctuations,
we take the deterministic ordered single-particle weights
\beq
q_m=1-\left(\frac{2m-1}{2M}\right)^{1/c}\qquad(m=1,\dots,M),
\label{qdet}
\eeq
corresponding to the most uniform non-random sampling of the
distribution~(\ref{fdef}).

\begin{figure}[!ht]
\begin{center}
\includegraphics[angle=-90,width=.4\linewidth]{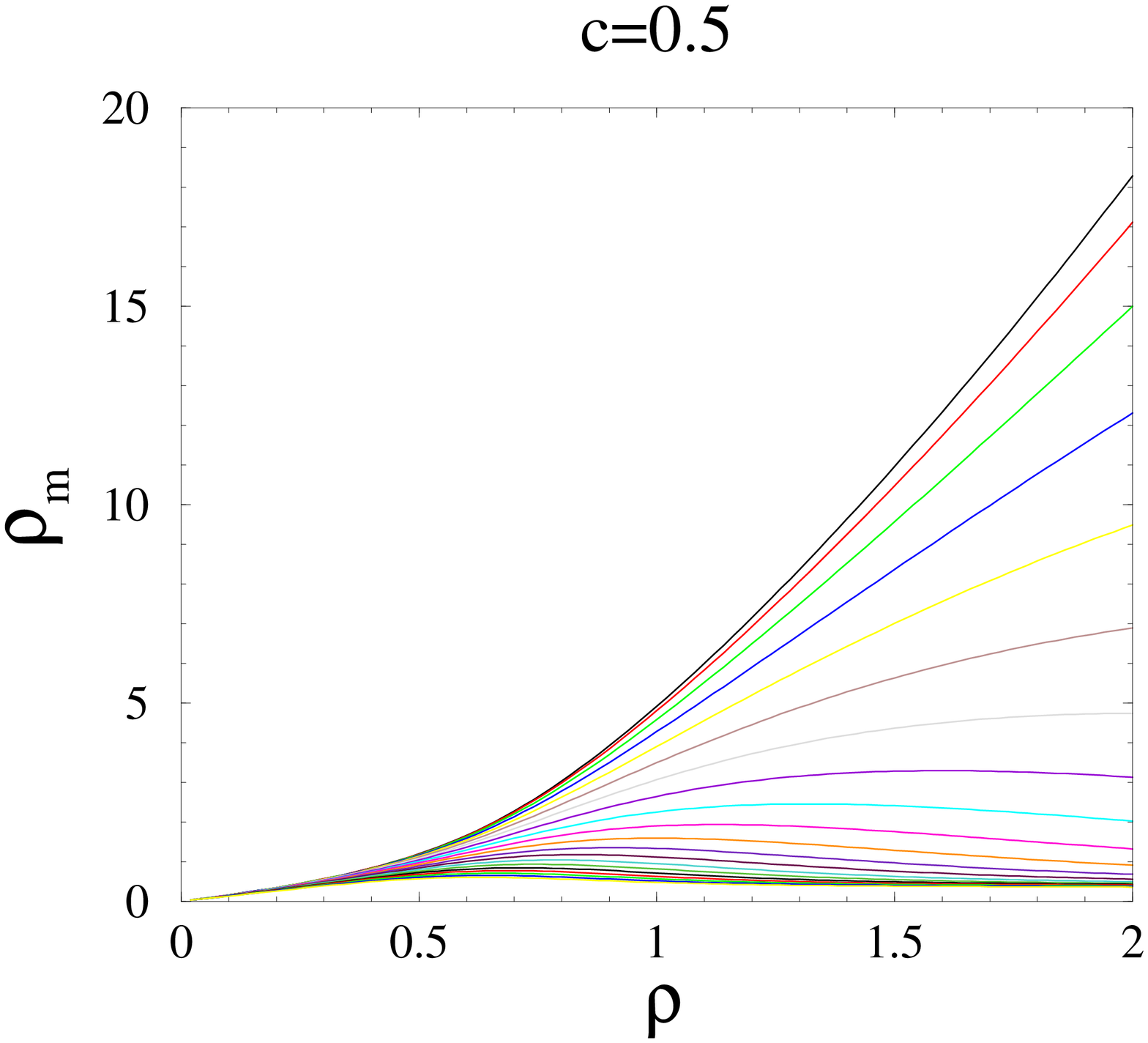}
{\hskip 8pt}
\includegraphics[angle=-90,width=.4\linewidth]{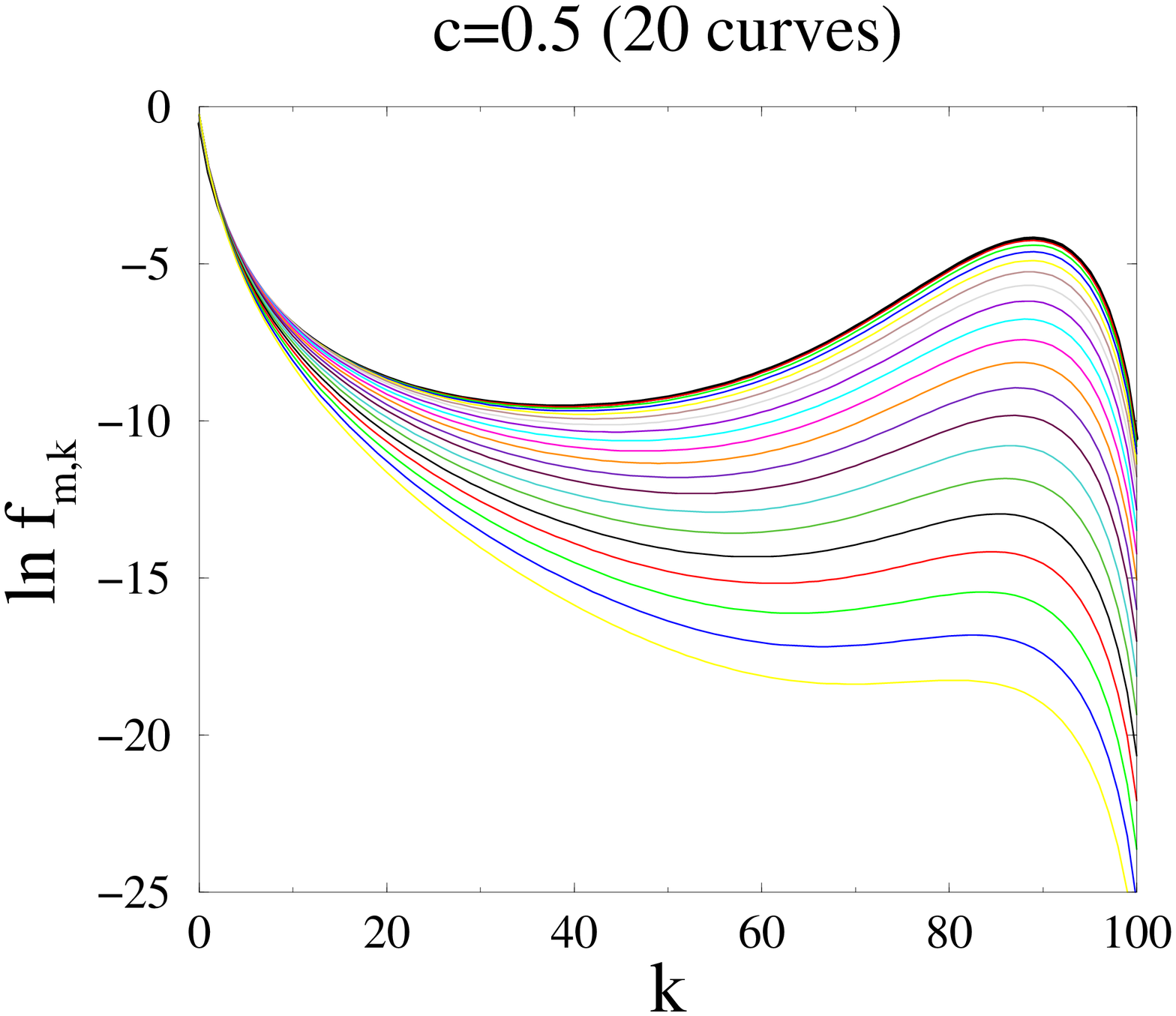}

\includegraphics[angle=-90,width=.4\linewidth]{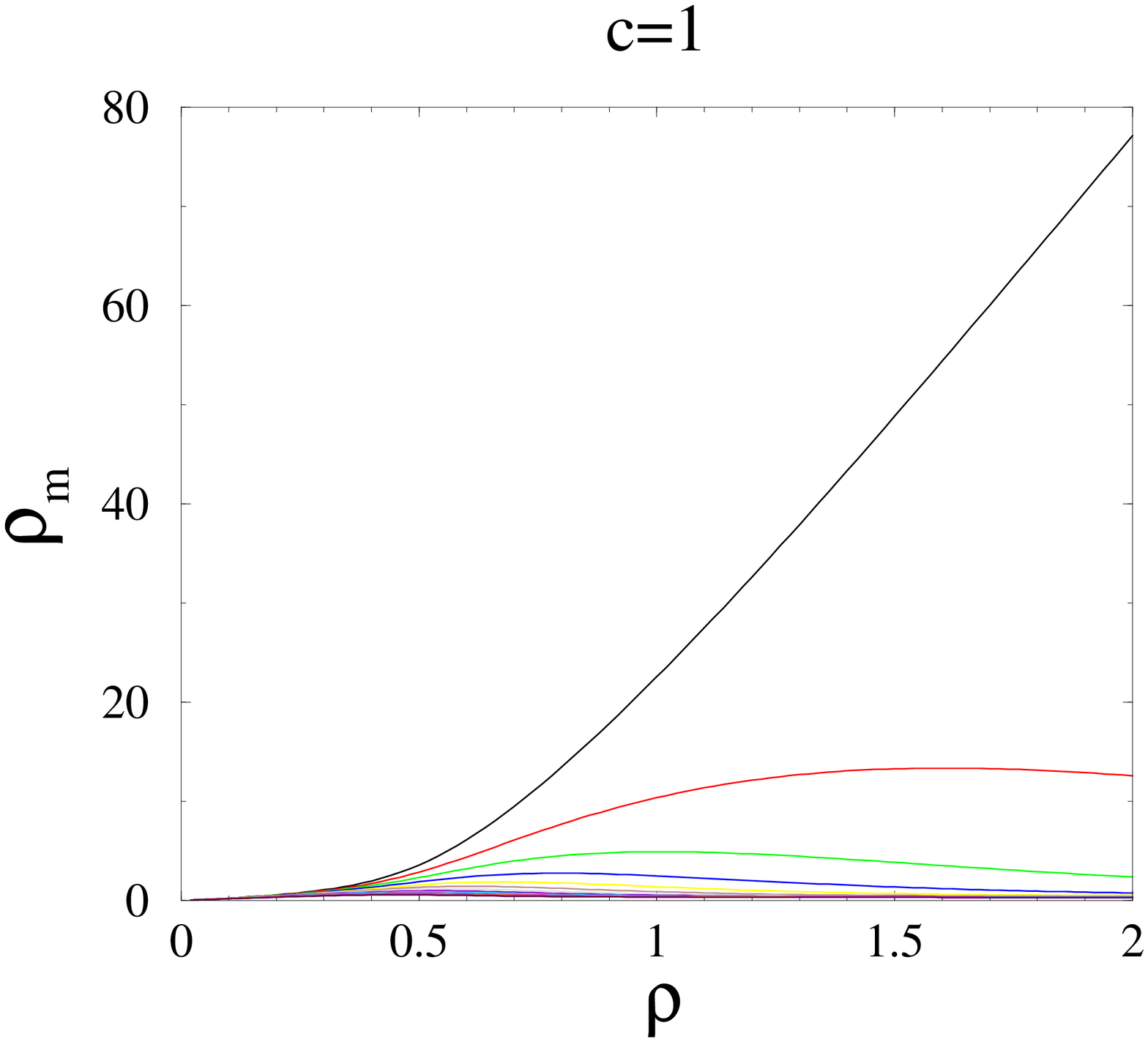}
{\hskip 8pt}
\includegraphics[angle=-90,width=.4\linewidth]{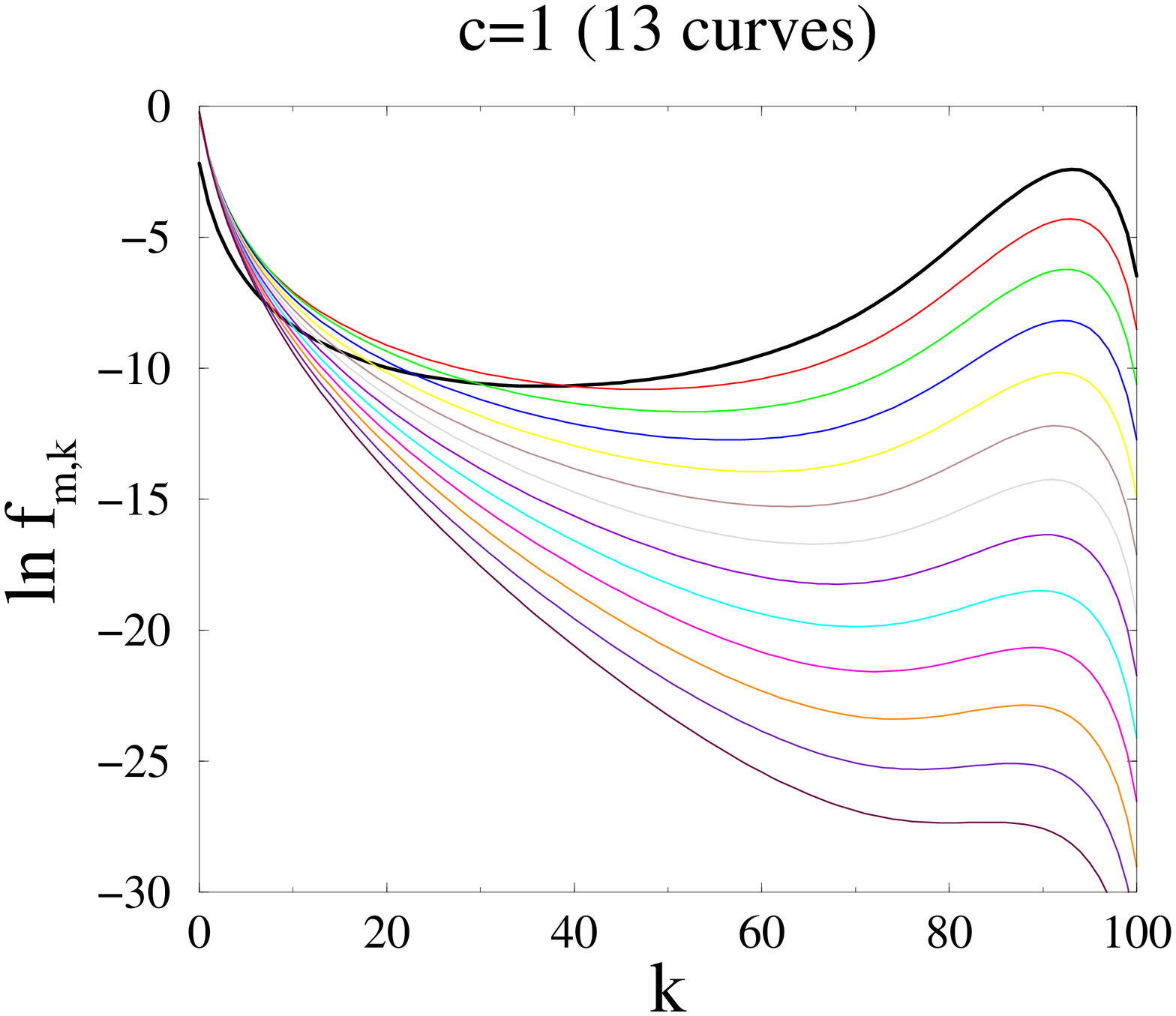}

\includegraphics[angle=-90,width=.4\linewidth]{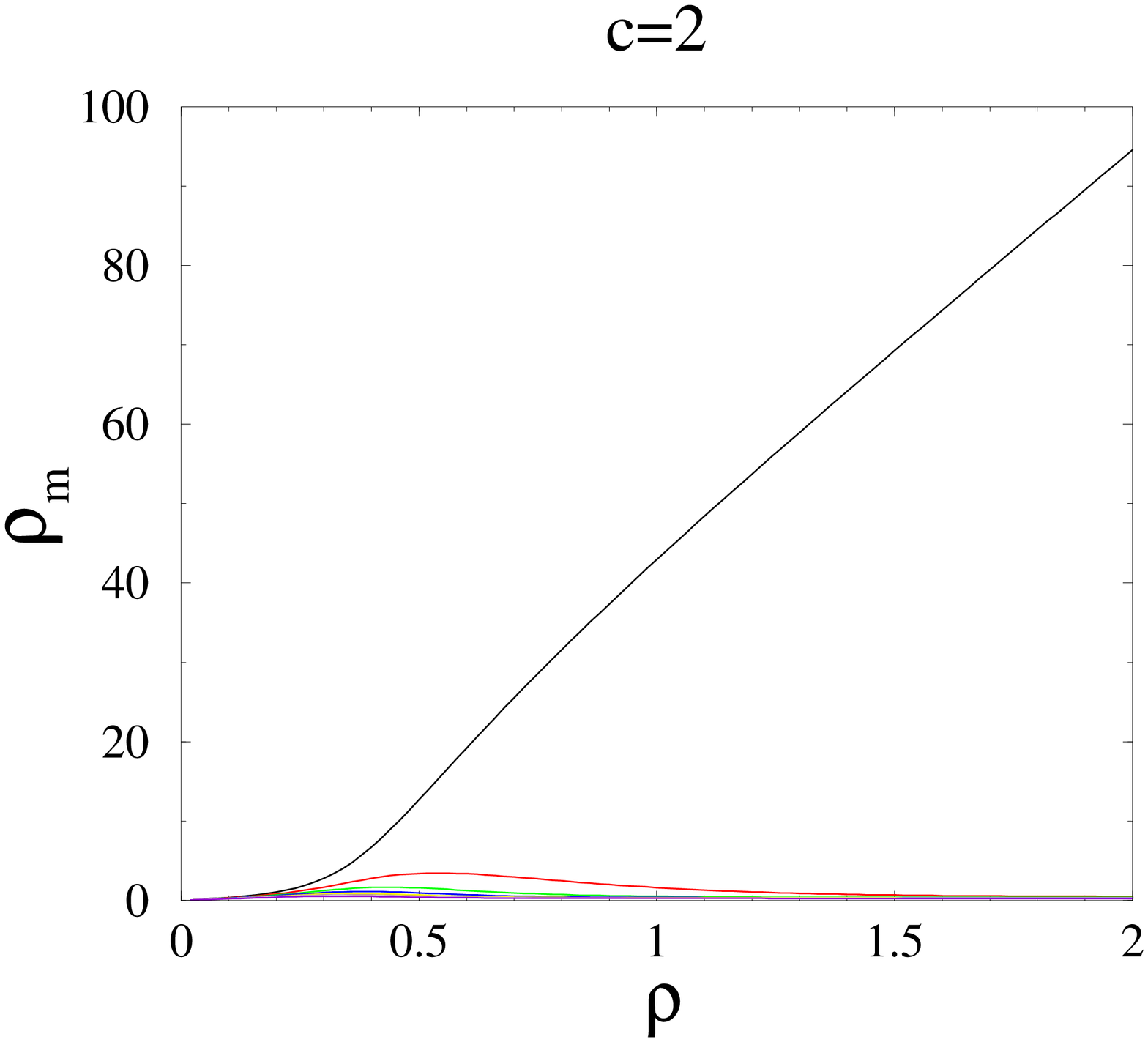}
{\hskip 8pt}
\includegraphics[angle=-90,width=.4\linewidth]{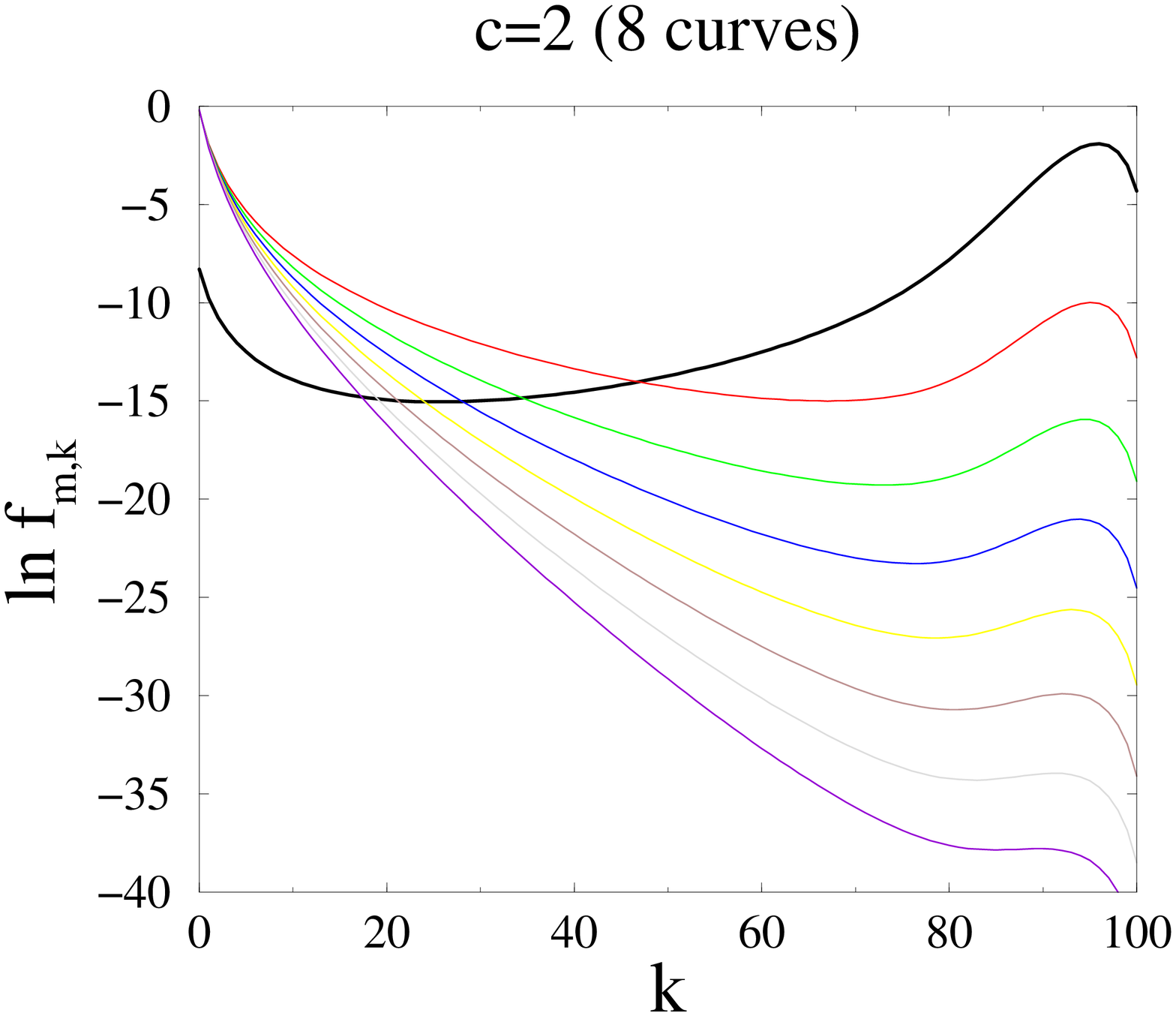}
\caption{\small
Left: plots of the local densities $\rho_m=\mean{N_m}$ at sites
$m=1,2,\dots$ against the mean density $\rho=N/M$,
for $M=50$ sites and up to $N_\max=100$ particles, with $b=4$.
Each panel corresponds to a value of $c$.
Each color consistently corresponds to a site label
(black for $m=1$, red for $m=2$, and so on).
Right: logarithmic plots of the occupation probabilities $f_{m,k}$
of the favored sites (see text) against particle number $k$,
for $M=50$ and $N=100$.
Other parameters and colors are the same as for the left-hand panels.}
\label{occ}
\end{center}
\end{figure}

The left-hand panels of figure~\ref{occ}
show plots of the canonical densities $\rho_m=\mean{N_m}$
of the successive sites ($m=1$, 2, $\dots$),
against the mean density $\rho=N/M$ of the system,
for $M=50$ and $N$ up to $N_\max=100$, i.e., $\rho_\max=2$,
for a fixed interaction exponent $b=4$,
and the deterministic single-particle weights~(\ref{qdet})
with three values of $c$.
The densities $\rho_m$ are calculated by means
of~(\ref{zrec})--(\ref{rhores}).
The ordering $q_1>q_2>q_3>\dots$ of the weights
always reflects itself into the same ordering for the densities,
i.e., $\rho_1\hbox{(black)}>\rho_2\hbox{(red)}>\rho_3\hbox{(green)}>\dots$
The various phases of the system are expected to manifest themselves
as different asymptotic growth laws in various parts of the density profile.
For $c=0.5$, the densities of the first few sites are observed to grow at similar rates,
at least at low enough density.
For $c=2$, only the most favored site has a rapidly growing density,
whereas all the other densities remain microscopic.
Finally, the situation for $c=1$ looks intermediate.
These findings are in agreement with the predicted transition at $c=1$
between an extended and a localized condensed phase.
Another feature is worth being noticed.
As a consequence of the interactions,
all the densities $\rho_m$ except the first one ($\rho_1$)
present a maximum for intermediate densities
and they eventually decrease at very high mean densities.
This phenomenon takes place for any positive values of $b$ and $c$.
It is of the same nature as the overshoot observed in the impurity
problem~\cite{angel}.

The right-hand panels of figure~\ref{occ} show logarithmic plots
of the local occupation probabilities $f_{m,k}$ of the successive sites
($m=1$, 2, $\dots$),
against $k$, for the same model parameters with $N=100$ particles on $M=50$ sites.
These occupation probabilities are calculated by means of~(\ref{fres}).
The favored sites, corresponding to the smaller values of $m$,
present a non-monotonic profile of occupation probabilities,
with a local minimum at $k_\min\approx\Delta/2$,
and a secondary peak at $k_\max\approx\Delta$, representing the condensate.
Only the data for those favored sites have been plotted for clarity.
The numbers of curves (20, 13, and 8)
therefore give the numbers of those favored sites.
For a given favored site $m$, the area under the secondary peak,
\beq
\Pi_m=\sum_{k=k_\min}^Nf_{m,k},
\label{pidef}
\eeq
gives an operational estimate of the hosting probability of site $m$.
As $c$ increases, the number of favored sites decreases,
and the hosting probabilities $\Pi_m$ decrease more and more rapidly
(remember the vertical scale on the right-hand panels of figure~\ref{occ} is logarithmic).

\begin{figure}[!ht]
\begin{center}
\includegraphics[angle=-90,width=.425\linewidth]{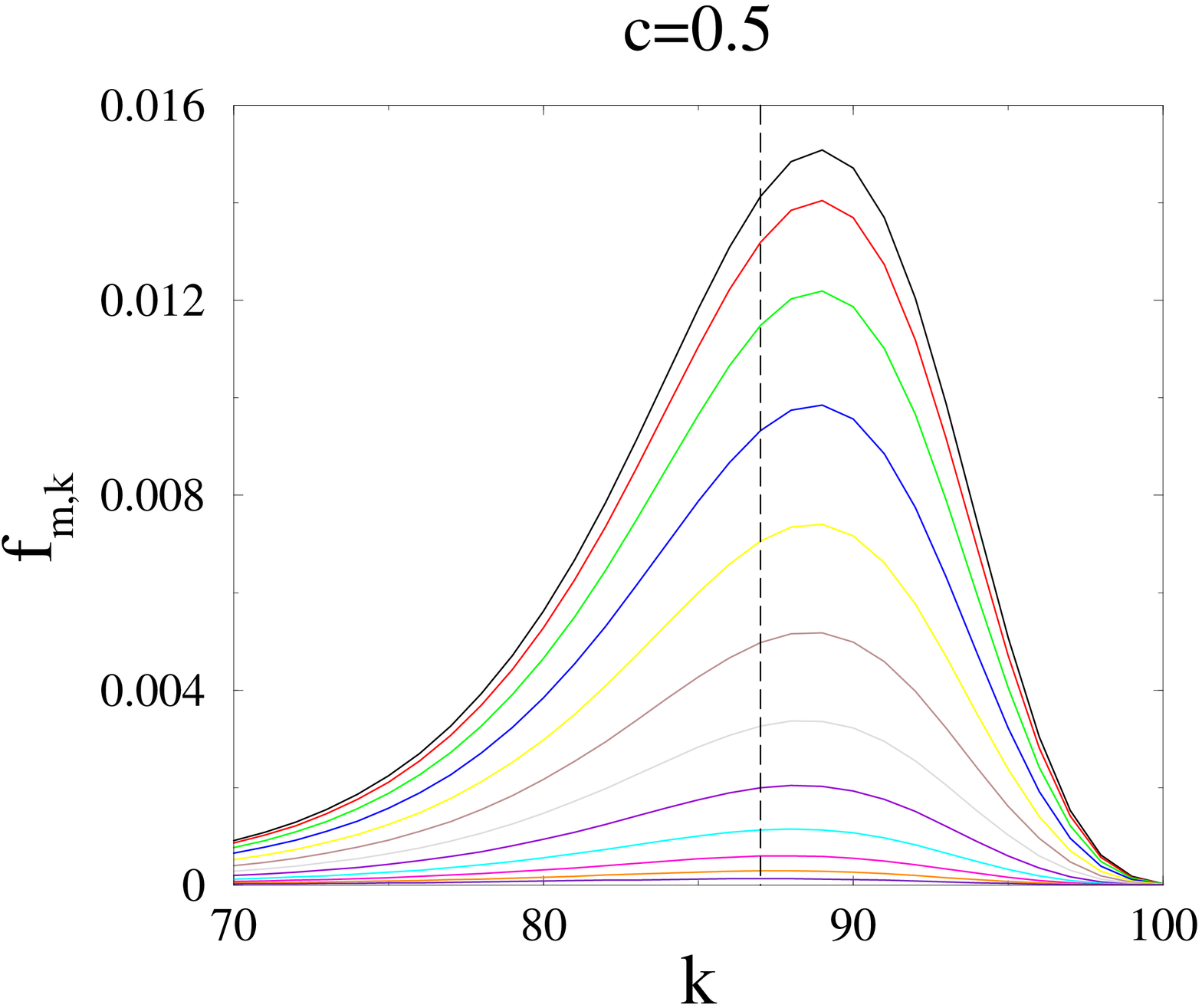}
{\hskip 8pt}
\includegraphics[angle=-90,width=.425\linewidth]{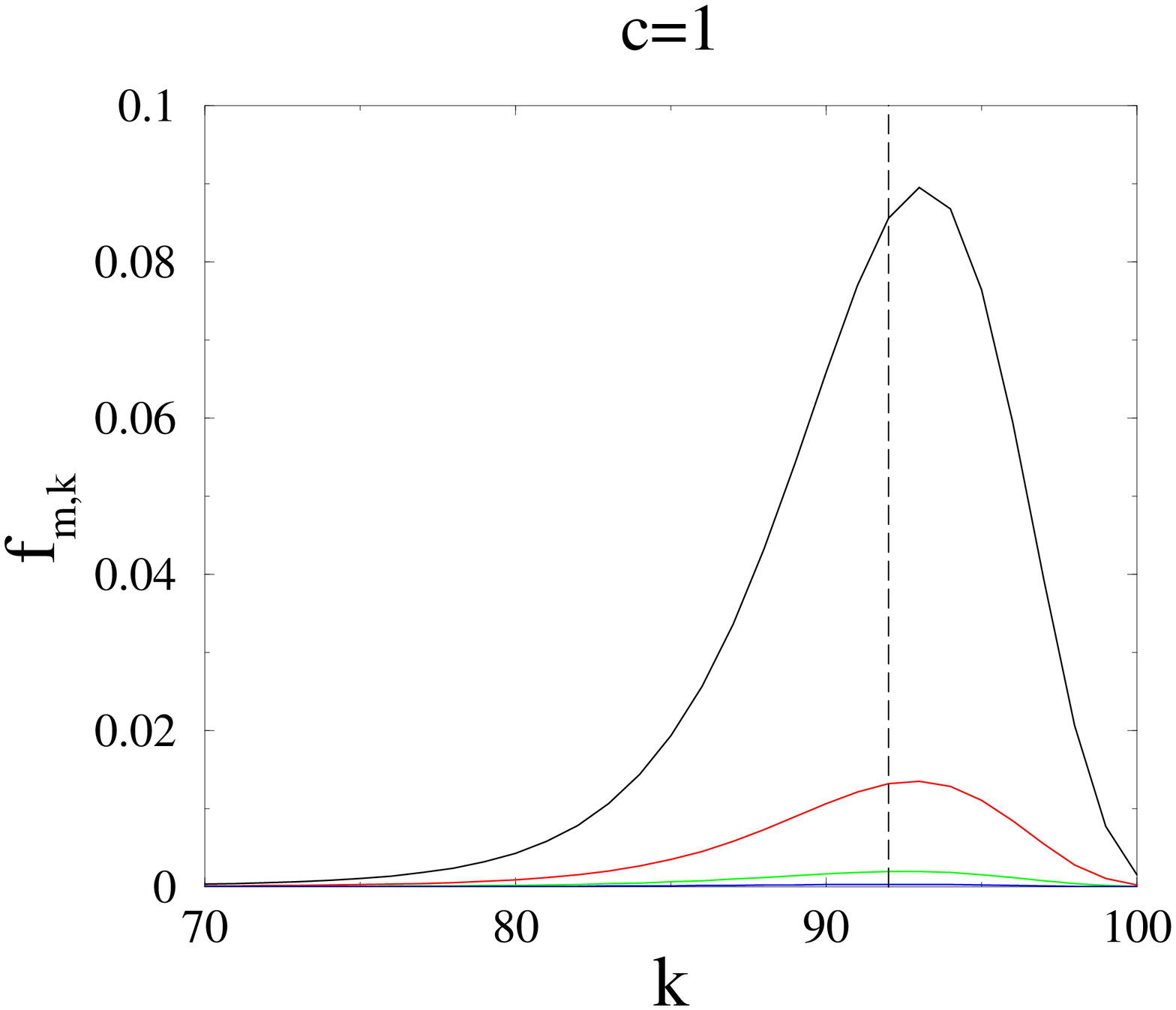}
\caption{\small
Enlargement of first two datasets presented on the right-hand side of figure~\ref{occ},
in the region of the condensate peak and on a linear scale.
Vertical dashed lines: thermodynamic predictions for the mean condensate size $\Delta$.}
\label{occlin}
\end{center}
\end{figure}

The role of the disorder exponent $c$ is further emphasized in figure~\ref{occlin},
showing an enlargement of the first two datasets presented
on the right-hand side of figure~\ref{occ},
in the region of the condensate peak and on a linear scale.
The contrast between the two situations corroborates our prediction.
For $c=0.5$ (in the extended phase),
a large number of favored sites exhibit a visible secondary peak,
and thus host the condensate with an appreciable probability.
For $c=1$ (at the borderline between the extended and localized phases),
only a small number of sites (4 on the plot) exhibit a visible secondary peak,
and the areas under the peaks decrease very rapidly.
In both cases the maxima of the peaks roughly coincide with the predictions
$k=\Delta=N-M\rho_c$, i.e., $k=87$ for $c=0.5$ ($\rho_c=0.2555$)
and $k=92$ for $c=1$ ($\rho_c=0.1641$) (vertical dashed lines).

\subsection{A quantitative study of the hosting probabilities}

We now turn to the most novel prediction of this work,
namely the existence of an extended condensed phase,
where the condensate can be hosted by a large set of favored sites,
whose size $R$ obeys the sub-extensive growth law~(\ref{kres}).

\begin{figure}[!ht]
\begin{center}
\includegraphics[angle=-90,width=.5\linewidth]{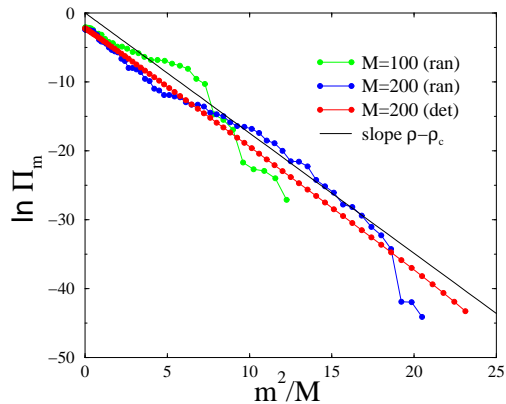}
\caption{\small
Logarithmic plot of the hosting probabilities $\Pi_m$
against $m^2/M$, for $b=4$, $c=1/2$ and $\rho=2$ in three cases (see text).
Black line with slope $\rho-\rho_c=1.7445$: theoretical estimate~(\ref{pig}).}
\label{w4}
\end{center}
\end{figure}

Let us first examine the individual hosting probabilities $\Pi_m$.
Considering $c=0.5$ for definiteness,
the estimate~(\ref{piasy}) simplifies to the half-Gaussian law
\beq
\Pi_m\sim\exp\left(-\frac{(\rho-\rho_c)m^2}{M}\right).
\label{pig}
\eeq
Figure~\ref{w4} shows a logarithmic plot of the hosting probabilities,
calculated by means of~(\ref{zrec}),~(\ref{fres}), and~(\ref{pidef}),
for all the sites $m$ having a condensate peak,
against $m^2/M$, for $b=4$ and $\rho=2$, in three situations:
a random sample of $M=100$ sites,
a random sample of $M=200$ sites,
and a deterministic sample of $M=200$ sites with single-particle
weights~(\ref{qdet}).
The black line with slope $\rho-\rho_c=1.7445$
shows the theoretical estimate~(\ref{pig}).
The dataset of the deterministic sample is found to accurately follow
the theoretical slope.
The data for the two random samples exhibit the expected trend,
although they are too noisy to be conclusive.
This observation justifies our choice to use deterministic single-particle weights
as illustrative examples in figures~\ref{occ} and~\ref{occlin}.
Finally, it is worth noticing that the sum of all the hosting probabilities,
\beq
S_1=\sum_{m=1}^M\Pi_m,
\eeq
is very close to unity in the three situations shown in figure~\ref{w4}.
We indeed respectively find $S_1-1=7.92\times10^{-6}$,
$4.73\times10^{-6}$, and $4.42\times10^{-6}$.
The asymptotic sum rule $S_1=1$, testifying the uniqueness of the condensate
in the thermodynamic limit,
is therefore verified with high accuracy for system sizes as small as $M=100$.

We proceed by an investigation of the size $R$ of the hosting set for
the condensate.
Along the lines of the theory of Anderson localization~\cite{ipr},
we can expect that a reliable estimate of $R$
will be provided by the inverse participation ratio,
\beq
R=\frad{\left(\sum_{m=1}^M\Pi_m\right)^2}{\sum_{m=1}^M\Pi_m^2},
\label{kpidef}
\eeq
i.e., the inverse of the quantity $Y$ defined in~(\ref{ydef}).
This operational definition holds for any finite sample.
The hosting probability $\Pi_m$ of a favored site $m$
is measured as the area under the condensate peak (see~(\ref{pidef})),
and set to zero for the unfavored sites where the occupation probabilities $f_{m,k}$
exhibit no such peak.
The normalization in the numerator of~(\ref{kpidef}) is needed for the smaller systems,
where the $\Pi_m$ do not exactly sum up to unity.
We now have a tool to help us revisiting the various regimes of the phase diagram.

\begin{itemize}

\item
{\it Extended regime ($c<1$).}
In this regime, it is legitimate to make use of the expression~(\ref{piasy})
for the hosting probabilities $\Pi_m$
in order to estimate the quantity~$R$ defined in~(\ref{kpidef}).
Evaluating the sums as integrals,
we end up with the result that $R$ is asymptotically self-averaging and grows as
\beq
R\approx A\,M^{1-c},\qquad
A=\left(\frac{2}{\rho-\rho_c}\right)^c\Gamma(c+1).
\label{kave}
\eeq
The scaling estimate~(\ref{kres}) has thus been turned into a quantitative prediction.
Figure~\ref{k575} shows plots of the mean size $\overline R$
of the hosting set, for $\rho=2$ and $b=3$ and $b=4$,
against $M^{1/2}$ for $c=0.5$ (left)
and against $M^{1/4}$ for $c=0.75$ (right).
All the data presented in figures~\ref{k575}--\ref{k2}
have been obtained by generating many (typically $10^5$)
independent configurations of $M$ single-particle weights $\{q_m\}$
drawn from the distribution~(\ref{fdef}),
and by exactly evaluating for each configuration the quantity~$R$
by means of~(\ref{zrec}),~(\ref{fres}),~(\ref{pidef}), and~(\ref{kpidef}).
All the datasets shown in figure~\ref{k575} exhibit a linear behavior,
corroborating the power-law growth~(\ref{kave}),
and show a reasonable quantitative agreement with the predicted amplitudes $A$.

\begin{figure}[!ht]
\begin{center}
\includegraphics[angle=-90,width=.425\linewidth]{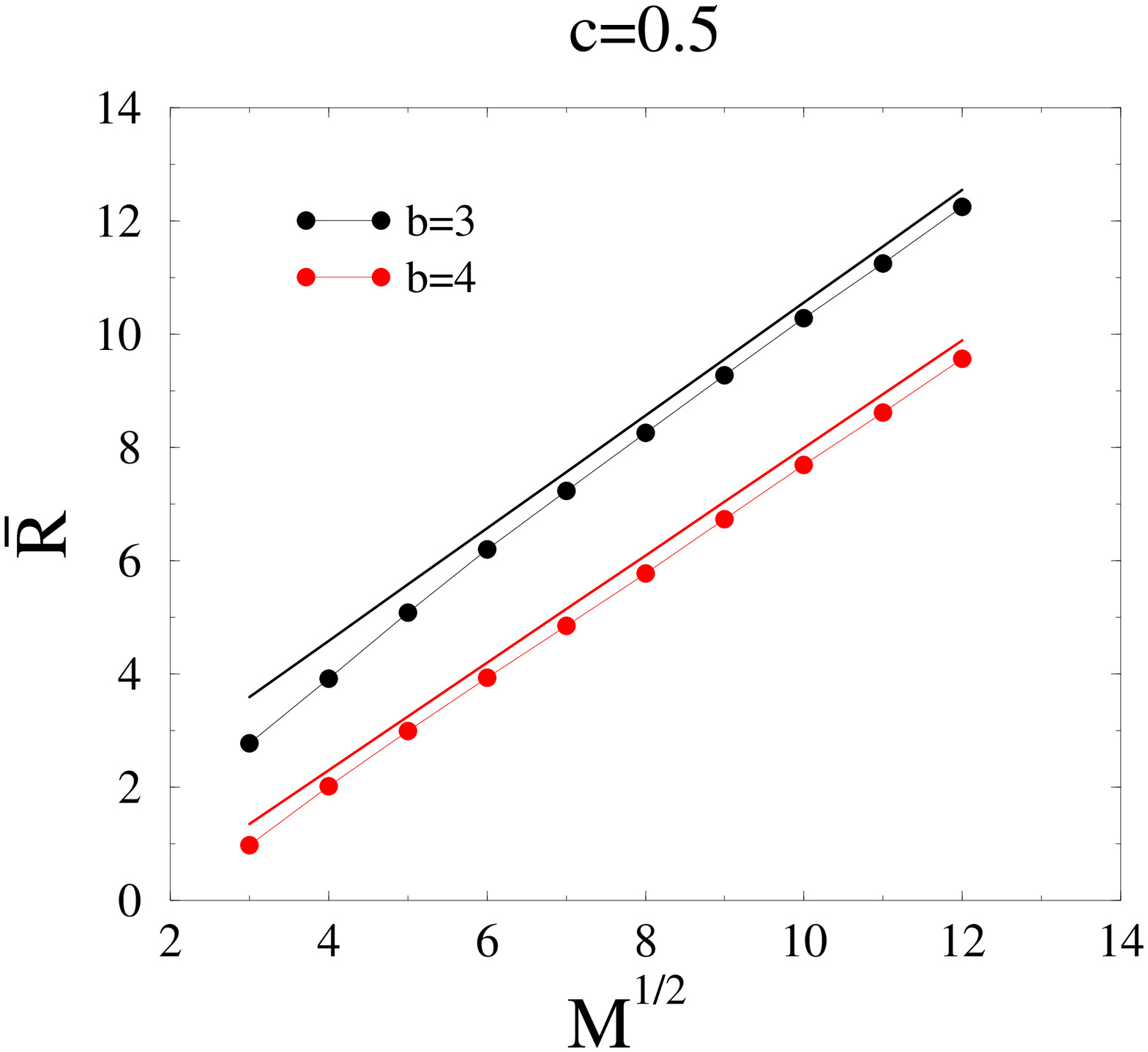}
{\hskip 8pt}
\includegraphics[angle=-90,width=.425\linewidth]{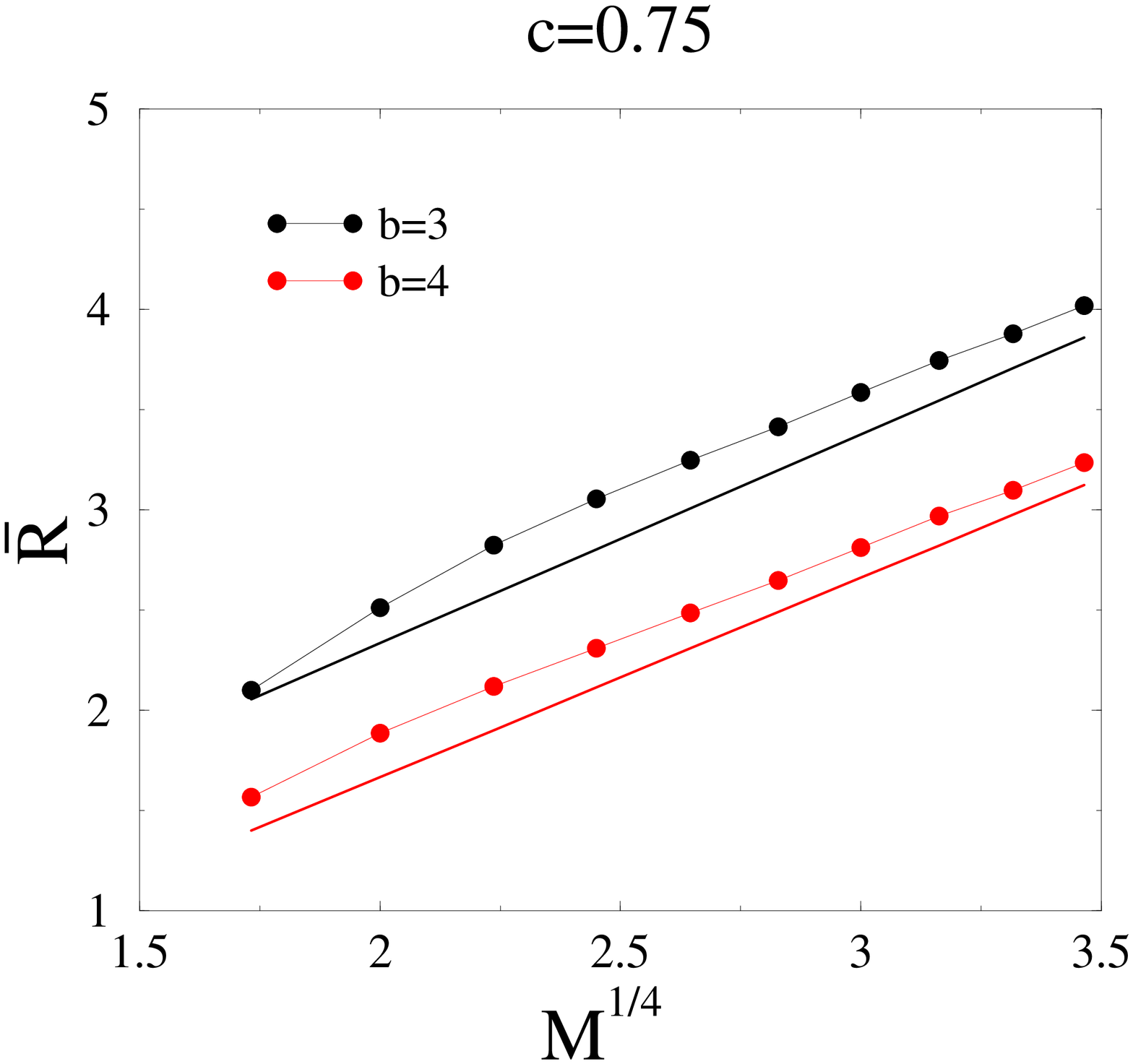}
\caption{\small
Mean size $\overline R$ of the hosting set in the extended regime.
Straight lines have the theoretical slopes $A$ (see~(\ref{kave})).
Left: Data for $c=0.5$ and $\rho=2$ against $M^{1/2}$.
For $b=3$, $\rho_c=0.4140$ and $A=0.9952$.
For $b=4$ (data shifted vertically for readability), $\rho_c=0.2555$ and $A=0.9489$.
Right: Data for $c=0.75$ and $\rho=2$ against $M^{1/4}$.
For $b=3$, $\rho_c=0.3078$ and $A=1.0418$.
For $b=4$ (data shifted vertically for readability), $\rho_c=0.2007$ and $A=0.9949$.}
\label{k575}
\end{center}
\end{figure}

\item
{\it Borderline case ($c=1$).}
This borderline situation is investigated in detail in~\ref{borderline}.
The quantity $R$, which identifies with $1/Y$ (see~(\ref{ydef})),
has a non-trivial limit distribution which only depends on $r=\rho-\rho_c$.
An expression for the mean value $\overline{R}$ is given in~(\ref{kapp}).
Figure~\ref{k1} shows a plot of $\overline{R}$ against $M$ for $\rho=2$ and
$b=3$ and $b=4$.
The data seem to exhibit a fast convergence to well-defined limits.
The apparent limiting values however show a slight disagreement
(of the order of half a percent in relative value)
with the theoretical predictions given in~(\ref{kapp})
(horizontal lines),
i.e., $\overline{R}=1.9411$ for $b=3$ ($\rho_c=0.2421$, $r=1.7579$),
and $\overline{R}=1.8995$ for $b=4$ ($\rho_c=0.1641$, $r=1.8359$).
The observed discrepancy is most certainly attributable to a very slow
convergence.
The data (not presented here) with the deterministic sampling~(\ref{qdet})
indeed exhibit a similar slow convergence, albeit toward different limiting values.

\begin{figure}[!ht]
\begin{center}
\includegraphics[angle=-90,width=.5\linewidth]{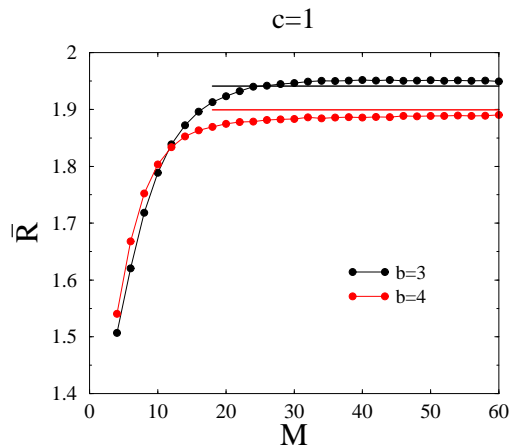}
\caption{\small
Mean size $\overline R$ of the hosting set
in the borderline case ($c=1$) against $M$,
for $\rho=2$ and $b=3$ and $b=4$.
Horizontal lines: predicted limiting values
$\overline{R}=1.9411$ for $b=3$ ($\rho_c=0.2421$, $r=1.7579$),
$\overline{R}=1.8995$ for $b=4$ ($\rho_c=0.1641$, $r=1.8359$) (see~(\ref{kapp})).}
\label{k1}
\end{center}
\end{figure}

\item
{\it Localized regime ($c>1$).}
In this regime, we argued in section~\ref{thermo}
that the condensate is localized on the most favored site
($m=1$) with very high probability,
hence $R$ should converge to unity for large systems.
The first correction to this trivial limit
comes from the rare events where the weights
$\Pi_1$ and $\Pi_2$ are comparable.
The probability for this to occur falls off as $1/M^{1-1/c}$.
In~\ref{effloc} this argument is turned
into the quantitative prediction (see~(\ref{appkloc}))
\beq
\overline{R}-1\approx\frac{\pi
c\,\Gamma(2-1/c)}{2(\rho-\rho_c)}\,M^{-(1-1/c)}.
\label{kloc}
\eeq
For $c=2$ this yields
\beq
\overline{R}-1\approx\frac{\pi^{3/2}}{2(\rho-\rho_c)\sqrt{M}}.
\label{kloc2}
\eeq
Figure~\ref{k2} shows a plot of $\overline{R}$ against $M$ for $b=4$ and $c=2$
(and so $\rho_c=0.0927$), and $\rho=2$.
The data exhibit a smooth maximum near $\overline{R}\approx1.39$ for $M=7$,
whereas they slowly go to unity for large systems,
in good agreement with the theoretical prediction~(\ref{kloc2}),
i.e., $\overline{R}\approx1+1.4597/\sqrt{M}$.

\begin{figure}[!ht]
\begin{center}
\includegraphics[angle=-90,width=.5\linewidth]{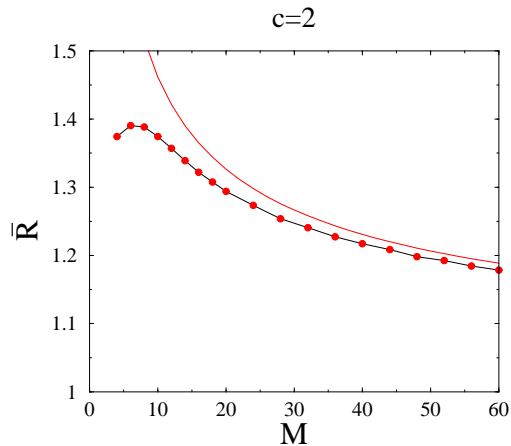}
\caption{\small
Mean size $\overline R$ of the hosting set
in the localized regime ($c=2$)
against $M$ for $b=4$ (and so $\rho_c=0.0927$), and $\rho=2$.
Full red line: theoretical prediction~(\ref{kloc2}),
i.e., $\overline{R}\approx1+1.4597/\sqrt{M}$.}
\label{k2}
\end{center}
\end{figure}

\end{itemize}

\section{Complex networks}
\label{networks}

In this section we investigate to what extent the above results
concerning localization properties of the condensate
apply to the ZRP on complex networks~\cite{nets}.
These networks are usually scale-free,
i.e., characterized by a broad distribution
of the node degrees $K_m$, falling off as the power law
$f_K\sim K^{-\gamma}$, with $\gamma>2$.

In the simplest situation where individual particles perform ordinary random walk,
the stationary weight $q_m$ at node $m$ is proportional to the degree $K_m$.
The ZRP with this underlying inhomogeneous diffusion has been investigated
by several groups~\cite{noh,krakow,tlz}.
Strictly speaking, this situation does not fit within the present work,
as node degrees are unbounded.
In other words, we have $q_\max=\infty$ and $z_c=0$,
and so the model does not have a fluid grand-canonical phase.
It has indeed been shown~\cite{noh} that this inhomogeneous ZRP
exhibits complete condensation (at least when $u_k=1$),
in the (weak) sense that the critical density $\rho_c$ vanishes in the thermodynamic limit.

Consider a large but finite network consisting of $M$ nodes.
We have thus approximately $\Delta\approx N=M\rho$.
The nodes which can host the condensate are those with highest degrees,
whose statistics has been studied recently~\cite{kr,ggl}.
The node whose degree $K_1$ is the highest is called the leader in those references.
The estimate $K_1\sim M^{1/(\gamma-1)}$ is easily obtained from an
argument of extreme-value statistics.
Two situations have to be dealt with separately.

\begin{itemize}

\item
There is a single {\it leader}, i.e., a single node with highest degree $K_1$.
In that case, we have $K_2=K_1-j$ for some $j=1,2,\dots$
and so $-\ln(\Pi_2/\Pi_1)\approx j\Delta/K_1\sim M^{(\gamma-2)/(\gamma-1)}$.
The hosting probability $\Pi_2$ is therefore exponentially small.
The condensate is therefore hosted by the unique leader with an overwhelmingly high probability.

\item
There are more than one node with highest degree $K_1$.
Let $C=2$, 3, $\dots$ denote the number of these {\it co-leaders}.
In that case, the condensate is hosted by each co-leader with equal probability,
and so we have $R=C$.
It has been shown in~\cite{ggl}
that the probability of having $C$ co-leaders decreases rapidly with $C$,
and that the leading event, namely the presence of two co-leaders ($C=2$),
occurs with a probability of the order of $1/K_1$.
We thus obtain the estimate
\beq
\overline{R}-1\sim M^{-1/(\gamma-1)}.
\eeq

\end{itemize}

There is a striking similarity between this result and
the prediction~(\ref{kloc}) in the localized regime ($c>1$).
The ZRP with inhomogeneous diffusion on complex networks is therefore always in the localized regime.
The quantitative correspondence between exponents reads
\beq
c=\frac{\gamma-1}{\gamma-2},\qquad\mbox{i.e.,}\quad\gamma=\frac{2c-1}{c-1}.
\eeq
The predictions made here are entirely due to the possible existence of co-leaders.

\section{Discussion}
\label{discussion}

In this work we have investigated the combined effects of interactions
and diffusion disorder on the condensation phenomenon in the inhomogeneous ZRP.
Our main findings have been summarized in the phase diagrams
shown in figures~\ref{phase1} and~\ref{phase2}.
These universal phase diagrams are drawn in the plane of two exponents,
the interaction exponent $b$ of the rate~(\ref{rate})
and the disorder exponent $c$ of the distribution~(\ref{fdef})
of the single-particle weights modeling inhomogeneous diffusion.

The most prominent feature put forward in this work is the existence
of an extended condensed phase.
This novel phase,
which corresponds to the domain where $0<c<1$ and $b>2-c$,
therefore shows up as a result of the combined effects of strong enough interaction
and weak enough disorder.
In this phase, a typical high-density configuration
has a unique condensate on top of a critical background,
but the condensate may be located at any site of a large hosting set of favored sites,
whose size grows sub-extensively as $R\sim M^{1-c}$.
The extended condensed phase
can therefore be viewed as implementing a continuous interpolation
between the two symmetry breaking scenarios known so far,
namely SSB in the homogeneous ZRP
and ESB in the occupation-independent inhomogeneous case.

It is worth underlining the qualitative difference between
the extended condensed phase of the inhomogeneous ZRP emphasized in this work
and several alternative condensation scenarios in interacting particle systems
which have been put forward in the recent literature.
The homogeneous ZRP with non-conventional hopping rates~$u_k$,
exhibiting either a non-monotonic dependence on the occupation $k$
or a weak explicit dependence on the system size,
may have several `condensates' in its stationary state~\cite{sem,tt}.
These new scenarios include the possibility of having an extensive number
of `condensates' whose typical size is finite but parametrically large,
or a sub-extensive number of `meso-condensates',
each of them having typically a sub-extensive population.
Some mass transport models, where the hopping rate (generalizing $u_k$)
also depends on the occupation of neighboring sites,
have been shown to develop an extended condensate,
i.e., a high-density structure containing all the excess particles
and extending over a number of sites of order $M^{1/2}$~\cite{ehm}.
The non-universal properties of the extended condensate in
this class of models have been underlined in~\cite{wsjm}.
Another variant of these models in one dimension leads to a phenomenon
of explosive condensation,
giving rise to a moving condensate sweeping the system at an accelerated pace~\cite{we}.
The `target process', introduced in~\cite{target}, is dual to the ZRP,
in the sense that the hopping rate now depends on the occupation of the arrival site.
In the case of asymmetric dynamics in two dimensions and above,
the stationary state again contains a one-dimensional extended condensate,
aligned with the direction of the mean current.
The `inclusion process',
where particles are allowed to make non-local jumps over the lattice,
exhibits unusual types of complete condensation,
including a case where almost all the particles are condensed
on the right-most side of a finite one-dimensional chain~\cite{grv}.
Finally, a non-Markovian generalization of the ZRP,
where the motion of particles is governed by internal clocks
endowed with their own dynamics,
leads to a variety of non-equilibrium stationary states.
In the asymmetric one-dimensional situation,
the model may exhibit a localized condensate sitting on two neighboring sites
and moving ballistically~\cite{hms}.

The present work focussed on the stationary state of the ZRP with inhomogeneous diffusion~(\ref{fdef}) and interaction~(\ref{rate}).
The dynamical consequences of our findings are clearly the next question of interest.
For the case of homogeneous diffusion on a large but finite system,
the ergodic motion of the condensate in the stationary state
has been investigated by the present authors in~\cite{us}.
The leading mechanism turns out to be that all the excess particles sequentially
quit the condensate and progressively rebuild it at another random distant site.
The associated characteristic time scale can therefore be analyzed
by means of an effective two-site model,
with the key ingredient being the occupation probability profile $f_k$ of the latter model.
This ergodic time scale is found to grow with the system size as the power law
\beq
\tau\sim\frac{\Delta^{b+1}}{M}\sim M^b,
\label{tausca}
\eeq
i.e., faster than the diffusive scale $M^2$ in the thermodynamic limit
(as $b>2$), but not exponentially fast.
(In the special situation of symmetric dynamics in one dimension,
our prediction is increased to $\tau\sim\Delta^{b+1}$.)
The scaling prediction~(\ref{tausca}) has been alluded to or used in several situations
germane to the present one~\cite{krakow,sem,tt,we,target,grv,hms}.
It has also been corroborated by a recent rigorous analysis~\cite{landim},
albeit in the regime where~$N$ becomes large while $M$ is kept finite.
Coming back to the inhomogeneous ZRP studied in the present work,
the occupation probability profile on two sites
exhibits a universal `dip', scaling as $N^{-b}$,
irrespective of the inhomogeneity (see~(\ref{fmin})).
It can therefore be expected that the scaling law~(\ref{tausca})
will still hold all over the extended condensed phase of the model (for $b>2$),
where the condensate is allowed to live on a large hosting set of favored sites.
In the region of the phase diagram where $2-c<b<2$ the ergodic time should scale as the diffusive one.
Finally the coarsening process of the formation of the condensate
is yet another facet of the dynamics worth investigating.

\subsection*{Acknowledgments}

It is a pleasure to thank Stefan Grosskinsky for interesting discussions.

\appendix

\section{The case of two sites}
\label{sectwosites}

Let us consider the simplest non-trivial situation of the inhomogeneous ZRP on two sites.
The quantities of interest can be given explicit expressions in this situation,
while its behavior is already illustrative of many features of the model on larger systems.

The single-particle weights $q_1$ and $q_2$
and the hopping rates $w_{12}$ and $w_{21}$ are related by
$w_{12}q_1=w_{21}q_2$ (see~(\ref{master})).
We introduce an inhomogeneity parameter $\eps$
so that $q_1/q_2=w_{21}/w_{12}=\e^{2\eps}$,
and parametrize the single-particle weights as
\beq
q_1=\e^\eps,\qquad q_2=\e^{-\eps}.
\eeq
For definiteness, we set $\eps>0$, so that $q_1>q_2$, i.e., site 1 is favored.

\subsection{The occupation-independent case ($u_k=1$)}
\label{twofree}

Let us start with the occupation-independent case ($u_k=1$),
and consider the canonical ensemble where the total number $N$ of particles is fixed.

The partition function~(\ref{zcfree}) becomes
\beq
Z_{2,N}=\frac{q_1^{N+1}-q_2^{N+1}}{q_1-q_2}
=\frac{\sinh(N+1)\eps}{\sinh\eps}.
\eeq
Setting $N_1=k$ and $N_2=N-k$,
the occupation probabilities read
\beq
f_k=\prob{N_1=k}
=\frac{q_1^kq_2^{N-k}}{Z_{2,N}}=\frac{\sinh\eps\,\e^{(2k-N)\eps}}{\sinh(N+1)\eps}.
\eeq
The occupation probability profile is thus exponentially increasing
from the least probable event (site 1 is empty, i.e., $k=0$)
to the most probable one (site 2 is empty, i.e., $k=N$).
The ratio between the smallest and the largest of these probabilities reads
\beq
\frac{f_0}{f_N}=\left(\frac{q_2}{q_1}\right)^N=\e^{-2N\eps}.
\label{fratio}
\eeq

When the number of particles $N$ becomes large,
in order for the populations of the two sites to remain comparable,
one has to keep the model in the regime of a weak inhomogeneity,
where $\eps$ scales as $1/N$.
In this regime, many results assume scaling forms,
involving the rescaled inhomogeneity parameter
\beq
\theta=N\eps.
\label{thdef}
\eeq
For instance, the normalized occupation probability profile becomes
\beq
f_k\approx\frac{\eps\,\e^{(2x-1)\theta}}{\sinh\theta}\qquad(0<x=k/N<1)
\eeq
and, in particular, the probability ratio~(\ref{fratio}) becomes $\e^{-2\theta}$.
The density contrast
\beq
\delta\rho=\rho_1-\rho_2=\mean{N_1-N_2}=\frac{\partial}{\partial\eps}\ln Z_{2,N}
\eeq
scales as
\beq
\delta\rho\approx N\,\Phi_0(\theta),
\eeq
where the scaling function
\beq
\Phi_0(\theta)=\cotanh\theta-\frac{1}{\theta}
\label{phi0}
\eeq
starts increasing linearly as $\Phi_0(\theta)\approx\theta/3$ for small $\theta$,
and slowly saturates to unity as $\Phi_0(\theta)\approx1-1/\theta$ at large~$\theta$.

Let us briefly consider the grand-canonical case.
We have (see~(\ref{rhofree}))
\beq
\rho\GC_1=\frac{z\e^{\eps}}{1-z\e^{\eps}},\qquad
\rho\GC_2=\frac{z\e^{-\eps}}{1-z\e^{-\eps}}.
\eeq
The fugacity $z$ is determined by imposing the mean total occupation
$\mean{N\GC}=\rho\GC_1+\rho\GC_2$.
Skipping details, we again find that the density contrast
$\delta\rho\GC=\rho\GC_1-\rho\GC_2$
obeys a scaling law of the form~(\ref{phi0}),
with $\theta=\mean{N\GC}\eps$ and
\beq
\Phi\GC_0(\theta)=\frac{\theta}{1+\sqrt{\theta^2+1}}.
\label{phigc0}
\eeq
This scaling function is rather close to its canonical counterpart~(\ref{phi0}).
It starts increasing as $\Phi\GC_0(\theta)\approx\theta/2$
and slowly saturates to unity as $\Phi\GC_0(\theta)\approx1-1/\theta$.
Both functions will be plotted in figure~\ref{twosites}.

\subsection{The general case}
\label{twogal}

We now turn to the general case of the inhomogeneous ZRP on two sites
with the rate~(\ref{rate}),
considering only the canonical ensemble
and keeping the same notations as above.

The occupation probabilities read
\beq
f_k=\prob{N_1=k}
=\frac{\e^{(2k-N)\eps}p_kp_{N-k}}{Z_{2,N}},
\eeq
where the partition function is
\beq
Z_{2,N}=\sum_{k=0}^N\e^{(2k-N)\eps}p_kp_{N-k}.
\label{z2}
\eeq

In the presence of attractive interactions, i.e., for $b>0$,
the most probable event is still that site 2 is empty (i.e., $k=N$),
and the result~(\ref{fratio}) still holds.
More interestingly, however,
the occupation probability profile approximately reads
\beq
f_k\approx\frac{\Gamma(b+1)^2}{Z_{2,N}}\,\frac{\e^{(2k-N)\eps}}{k^b(N-k)^b},
\eeq
whenever both populations are large (i.e., $k\gg1$ and $N-k\gg1$).
The above expression is the product of an increasing exponential (reflecting the inhomogeneity)
and of two decaying power laws (reflecting the attractive interactions).

\begin{figure}[!ht]
\begin{center}
\includegraphics[angle=-90,width=.5\linewidth]{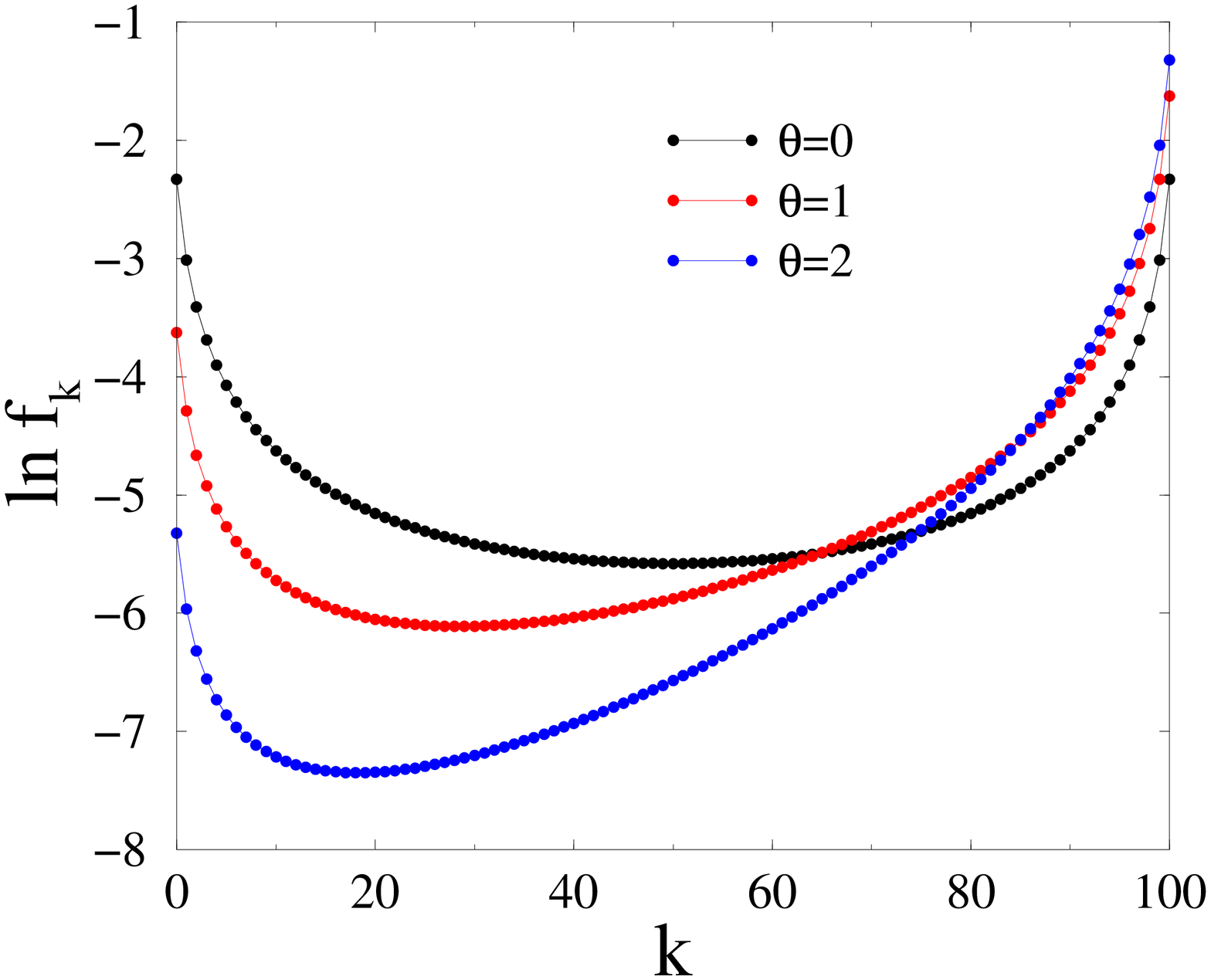}
\caption{\small
Logarithmic plot of the occupation probabilities $f_k$
for $N=100$ particles on two sites, with $b=1$ and several values of $\theta$.}
\label{occtwosites}
\end{center}
\end{figure}

In the scaling regime of a weak inhomogeneity
($N$ large, $\eps$ small, $\theta=N\eps$ fixed),
the above probability profile has a deep minimum
for a non-trivial value $k_\min$ of the occupation $k$, namely
\beq
k_\min\approx A(\theta)\,\frac{N}{2},
\label{kmin}
\eeq
with
\beq
A(\theta)=\frac{2b}{b+\theta+\sqrt{b^2+\theta^2}}.
\eeq
The corresponding smallest probability scales as
\beq
\frac{f_\min}{\sqrt{f_0f_N}}\approx B(\theta)\,2^{2b}\,\Gamma(b+1)\,N^{-b},
\label{fmin}
\eeq
with
\beq
B(\theta)=\left(\frac{b+\sqrt{b^2+\theta^2}}{2b}\right)^b.
\eeq
The probability profile therefore exhibits a universal {\it dip},
in the language of~\cite{us},
whose scaling in $N^{-b}$ holds irrespectively
of the rescaled inhomogeneity parameter~$\theta$,
which only enters the scaling functions $A(\theta)$ and $B(\theta)$.
These functions are normalized so that $A(0)=B(0)=1$ in the homogeneous limit.
The occurrence of a universal dip in the occupation probability profile
is the gist of the uniqueness of the condensate in the inhomogeneous ZRP for any $b>0$.
It also has far-reaching dynamical consequences,
which are briefly discussed in section~\ref{discussion}.
Figure~\ref{occtwosites} shows a logarithmic plot of the occupation probabilities $f_k$
for $N=100$ particles on two sites, for $b=1$ and several values of $\theta$.

In order to derive more quantitative predictions
concerning the scaling behavior of the partition function $Z_{2,N}$
and of the density contrast $\delta\rho$,
the following two regimes have to be considered separately,
along the lines of~\cite{us}.

\begin{itemize}

\item
For $b>1$, the sum in~(\ref{z2}) is dominated by configurations
such that almost all the particles
sit on one site, i.e., either $k$ is finite, or $N-k$ is finite.
Indeed the sum $P(1)$ of the~$p_k$ is convergent (see~(\ref{part})).
In other words, there is complete condensation in the $N\to\infty$ limit.

We thus obtain in the scaling regime
\beq
Z_{2,N}\approx\frac{2b\Gamma(b+1)}{b-1}\,N^{-b}\,\cosh\theta
\eeq
and
\beq
\delta\rho\approx N\,\Phi_1(\theta),
\eeq
where
\beq
\Phi_1(\theta)=\tanh\theta
\label{phi1}
\eeq
is independent of $b$ in the range $b>1$.
The above scaling function starts linearly as $\Phi_1(\theta)\approx\theta$
and saturates exponentially fast to unity as
$\Phi_1(\theta)\approx1-2\e^{-\theta}$.

\item
For $b<1$, the sum of the $p_k$ is divergent, and so all values of $k$ contribute
to the sum in~(\ref{z2}).
Evaluating this sum in the scaling regime
as an integral over $y=2x-1=2k/N-1$, we obtain
\beq
Z_{2,N}\approx\frac{(\pi b)^2}{\Gamma(2-2b)\,\sin^2(\pi b)}N^{1-2b}\,F_b(\theta),
\eeq
with
\beqa
F_b(\theta)&=&\frac{\Gamma(3/2-b)}{\sqrt{\pi}\,\Gamma(1-b)}
\int_{-1}^{+1}\frac{\e^{\theta y}\,\d y}{(1-y^2)^b}\nonumber\\
&=&\Gamma(3/2-b)(\theta/2)^{b-1/2}I_{1/2-b}(\theta),
\label{fbint}
\eeqa
where $I_\nu(\theta)$ denotes the modified Bessel function of index $\nu$.
The above scaling function is again normalized so that $F_b(0)=1$
in the homogeneous limit.
We thus have
\beq
\delta\rho\approx N\,\Phi_b(\theta),
\eeq
with
\beq
\Phi_b(\theta)=\frac{F'_b(\theta)}{F_b(\theta)}
=\frac{I_{3/2-b}(\theta)}{I_{1/2-b}(\theta)}.
\label{phib}
\eeq
This scaling function starts linearly as $\Phi_b(\theta)\approx\theta/(3-2b)$
and slowly saturates to unity as $\Phi_b(\theta)\approx1-(1-b)/\theta$.
The result~(\ref{phi0}) is recovered for $b=0$.

The function $\Phi_b(\theta)$
coincides with the function describing Langevin paramagnetism
in dimension $d=3-2b$.
Indeed, setting $y=\cos u$, the integral~(\ref{fbint}) can be recast as
\beq
F_b(\theta)=\frac{\Gamma(d/2)}{\sqrt{\pi}\,\Gamma((d-1)/2)}
\int_0^\pi(\sin u)^{d-2}\e^{\theta\cos u}\,\d u,
\eeq
where we recognize the normalized partition function for a classical spin,
represented by a vector ${\bf s}$ describing the unit sphere in $d$ dimensions,
coupled to a reduced magnetic field $\theta$,
so that $u$ is the angle between ${\bf s}$ and the direction of the field.
This analogy makes sense as long as ${\bf s}$ has continuous degrees of freedom,
namely for $d>1$, i.e., precisely $b<1$.

\end{itemize}

\begin{figure}[!ht]
\begin{center}
\includegraphics[angle=-90,width=.5\linewidth]{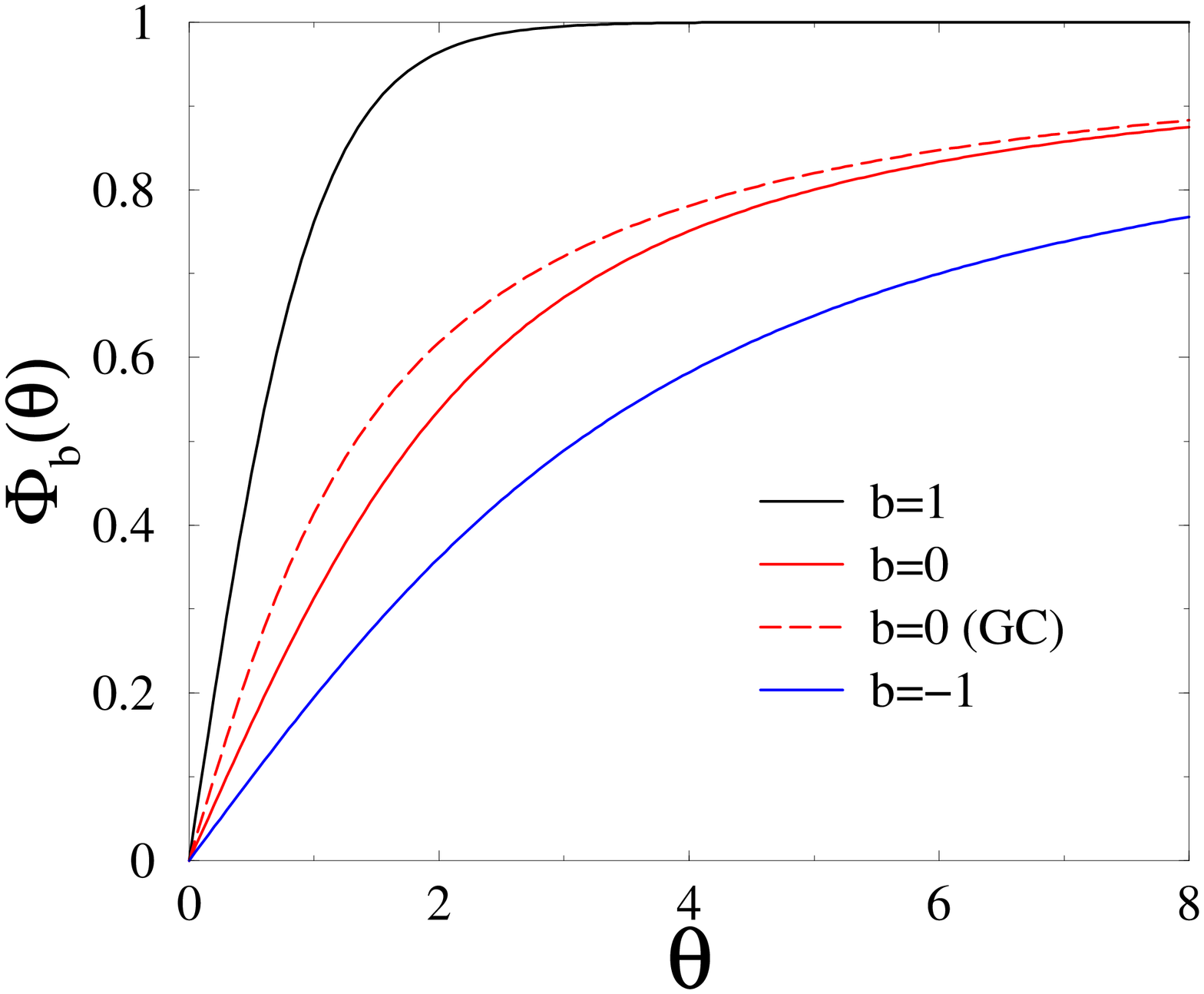}
\caption{\small
Finite-size scaling functions $\Phi_b(\theta)$
describing the density contrast of the two-site problem,
against the rescaled inhomogeneity parameter $\theta$,
for various values of the exponent $b$.
The grand-canonical scaling function
$\Phi\GC_0(\theta)$ is also plotted for comparison.}
\label{twosites}
\end{center}
\end{figure}

Figure~\ref{twosites} shows the
scaling function $\Phi_b$ of the density contrast for several $b$.
This function is independent of $b$ and saturates exponentially fast for $b>1$,
whereas it depends continuously on $b$ and saturates slowly for $b<1$.
Furthermore, at least for the case $b=0$, the grand-canonical scaling function
is very close to its canonical counterpart.

\section{Universal fluctuations in the borderline situation ($c=1$)}
\label{borderline}

This Appendix is devoted to a detailed study
of the fluctuations of the hosting probabilities $\Pi_m$
of a large system in the borderline situation ($c=1$).
We are thus led to consider an infinite ordered sequence
of unnormalized weights of the form (see~(\ref{pic}))
\beq
\Pi_m=\e^{-rx_m}\qquad(m=1,2,\dots),
\eeq
where the $x_m$ are Poissonian points with unit density,
whereas $r=\rho-\rho_c$ is the control parameter of the problem.

We are interested in the joint distribution of the two sums
\beq
S_1=\sum_{m\ge1}\Pi_m,\qquad
S_2=\sum_{m\ge1}\Pi_m^2,
\label{sumsdef}
\eeq
and chiefly in the distribution of the ratio
\beq
Y=\frac{S_2}{S_1^2},
\label{ydef}
\eeq
introduced in~\cite{df}.

The formula~(\ref{xrenewal}) allows us to recast the definition of the sums
$S_1$ and~$S_2$ into the form
\beqa
S_1&=&X_1+X_1X_2+X_1X_2X_3+\cdots,
\nonumber\\
S_2&=&X_1^2+X_1^2X_2^2+X_1^2X_2^2X_3^2+\cdots,
\label{ssum}
\eeqa
where the $X_m=\e^{-r\tau_m}$ are i.i.d.~random variables with the distribution
\beq
f_X(X)=\frac{1}{r}X^{-1+1/r}\qquad(0<X<1).
\label{fX}
\eeq
Infinite random sums such as~(\ref{ssum}) are known as Kesten
variables~\cite{kesten,vervaat}.
They have been shown to play a role in various areas
of the physics of one-dimensional disordered systems~\cite{dh,calan,theo}.
In the present situation, we have the identities
\beq
S_1\equiv X(1+S_1'),\qquad
S_2\equiv X^2(1+S_2'),
\label{kesiden}
\eeq
where $X$ is drawn from the distribution~(\ref{fX})
and $S_1'$ and $S_2'$ are copies of the variables $S_1$ and $S_2$,
independent of $X$.
The above identities can be viewed as a stochastic dynamical system
for two degrees of freedom, described by the coupled variables $S_1$ and $S_2$,
and submitted to the same noise $X$.

\subsection{Distribution of the normalization sum $S_1$}

The distribution of a Kesten variable such as $S_1$ can only be worked out
for special distributions of the $X_m$~\cite{calan}.
The present situation,
where the $X_m$ have a power-law distribution on the interval $[0, 1]$,
is the simplest of all the exactly solvable cases~\cite{vervaat}.
It is convenient to use the Laplace transform of the distribution,
\beq
L_1(t)=\overline{\e^{-tS_1}}.
\eeq
The identity~(\ref{kesiden}) for $S_1$ yields
\beq
L_1(t)=\overline{\e^{-tX(1+S_1')}}=\int_0^1 f_X(X)\,\e^{-tX}\,L_1(tX)\,\d X.
\eeq
The change of variable from $X$ to $u=tX$ leads to
\beq
L_1(t)=\frac{1}{r}\,t^{-1/r}\int_0^t u^{-1+1/r}\,\e^{-u}\,L_1(u)\,\d u,
\eeq
hence the differential equation
\beq
L_1'(t)=-\frac{1-\e^{-t}}{rt}\,L_1(t)
\eeq
and the explicit expression
\beq
L_1(t)=\e^{-F_1(t)/r},
\label{laps1}
\eeq
with
\beqa
F_1(t)&=&\int_0^1(1-\e^{-tx})\frac{\d x}{x}
\nonumber\\
&=&\int_0^t\frac{1-\e^{-u}}{u}\,\d u
=\ln t+\gamma-{\rm Ei}(-t)
=\sum_{n\ge1}\frac{(-1)^{n-1}t^n}{n\,n!},
\label{Fdef}
\eeqa
where $\gamma=0.577\,215\dots$ denotes Euler's constant,
and Ei is the exponential integral.

The series in the rightmost side of~(\ref{Fdef})
demonstrates that the cumulants of $S_1$ have the simple expression
\beq
\cum{S_1^n}=\frac{1}{nr}.
\eeq
We have in particular $\overline{S_1}=1/r$ and $\var S_1=1/(2r)$,
and so $1/r$ can be interpreted as the effective number of significant weights
which contribute to the sums~(\ref{sumsdef}).

The full distribution of~$S_1$ reads
\beq
f_{S_1}(S_1)=\int\frac{\d t}{2\pi\ii}\,\e^{tS_1-F_1(t)/r}.
\label{fcon}
\eeq

\begin{figure}[!ht]
\begin{center}
\includegraphics[angle=-90,width=.5\linewidth]{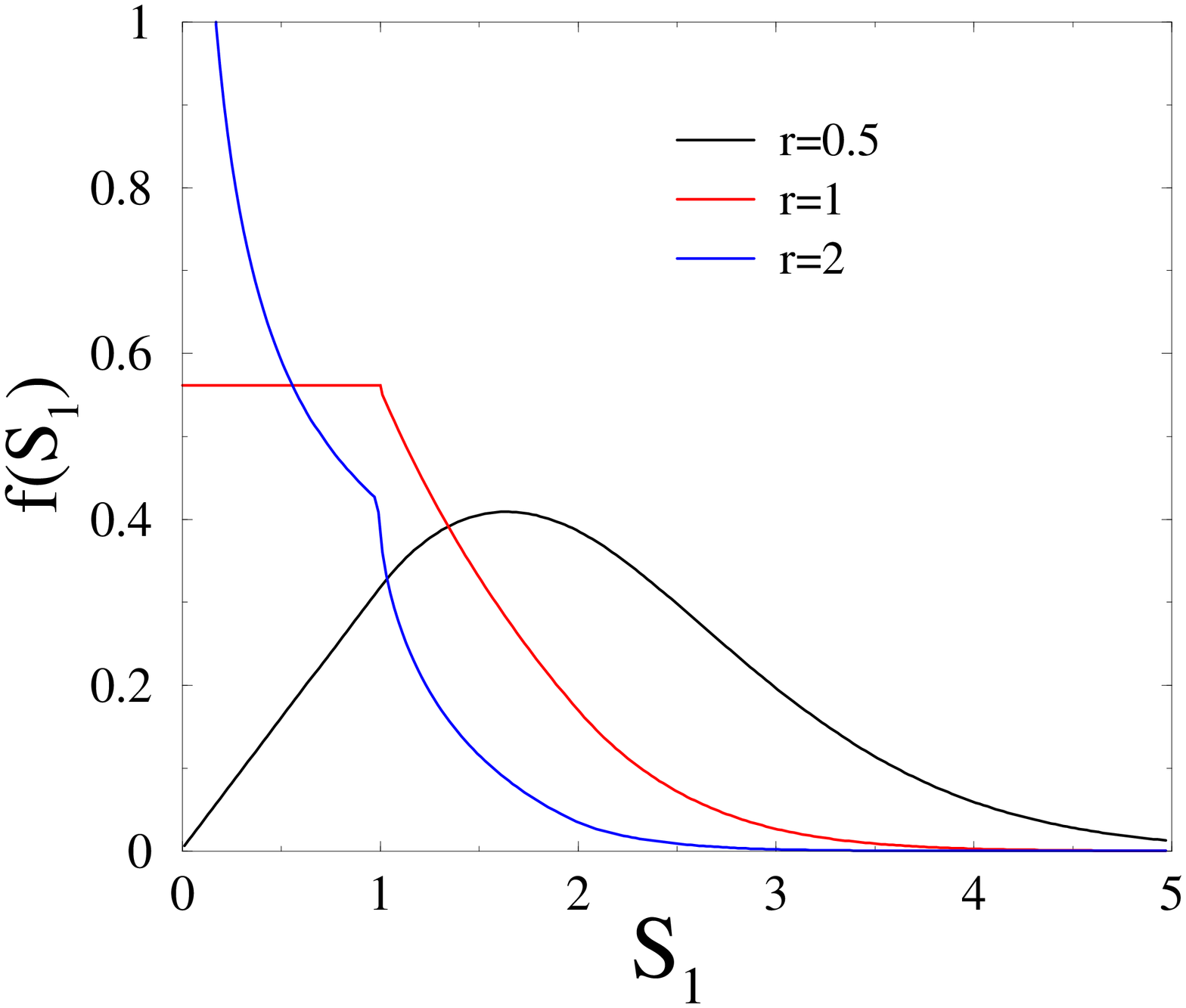}
\caption{\small
Distribution $f_{S_1}(S_1)$ of the normalization sum $S_1$,
for three characteristic values of the control parameter $r$.}
\label{shisto}
\end{center}
\end{figure}

Figure~\ref{shisto} shows a plot of this distribution
for three characteristic values of the control parameter $r$.
Each dataset consisting of 200 bins has been obtained by means of $10^9$
iterations of the random recursion~(\ref{kesiden}).

The distribution $f_{S_1}$ becomes a narrow Gaussian in the $r\to0$ limit,
in qualitative agreement with the law of large numbers,
as the number $1/r$ of significant weights is large.
It gets progressively broader as $r$ increases.

The behavior of $f_{S_1}$ for small values of $S_1$
can be estimated by using large positive values of $t$
in the contour integral~(\ref{fcon}).
Neglecting the exponentially small function ${\rm Ei}(-t)$,
we readily obtain the power~law
\beq
f_{S_1}(S_1)\approx\frac{\e^{-\gamma/r}}{\Gamma(1/r)}\,S_1^{-1+1/r}\qquad(S_1\to0).
\label{fsmall}
\eeq
The density therefore diverges as $S_1\to0$ for $r>1$,
whereas it tends to zero for $r<1$.
This feature is clearly visible on figure~\ref{shisto}:
$f_{S_1}$ vanishes linearly for $r=1/2$, goes to the constant $\e^{-\gamma}$
for $r=1$,
and diverge as $1/\sqrt{S_1}$ for $r=2$.

The behaviour of $f_{S_1}(S_1)$ for large values of $S_1$
can be estimated from the contour integral~(\ref{fcon})
by means of the saddle-point method.
Skipping details, we are left with a super-exponential tail of the form
\beq
f_{S_1}(S_1)\sim\exp\Bigl(-S_1(\ln S_1+\ln(\ln S_1)+\ln r+\cdots)\Bigr).
\eeq

Finally, performing an integration by parts in~(\ref{fcon}),
we obtain the following differential-difference equation
for the distribution $f_{S_1}(S_1)$:
\beq
(r-1)f_{S_1}(S_1)+rS_1f_{S_1}'(S_1)+f_{S_1}(S_1-1)=0.
\eeq
This functional equation has many consequences,
including that the estimate~(\ref{fsmall}) holds identically over the interval
$0<S_1<1$,
and that the distribution $f_{S_1}(S_1)$ has weaker and weaker singularities
at all the integer values $S_1=1,2,\dots$
Distributions of this kind have been met in several instances,
starting with R\'enyi's analysis of the one-dimensional car parking
problem~\cite{parking}.

\subsection{Joint distribution of the sums $S_1$ and $S_2$}

The joint distribution $f_{S_1,S_2}(S_1,S_2)$ of the sums $S_1$ and $S_2$
can be studied along the same lines.
Its Laplace transform reads
\beq
L_2(t_1,t_2)=\overline{\e^{-t_1S_1-t_2S_2}}.
\eeq
The identities~(\ref{kesiden}) yield
\beq
L_2(t_1,t_2)=\int_0^1 f_X(X)\,\e^{-t_1X-t_2X^2}\,L_2(t_1X,t_2X^2)\,\d X.
\eeq
Setting $g=t_2/t_1^2$, the change of variable from $X$ to $u=t_1X$ leads to
\beq
L_2(t_1,gt_1^2)
=\frac{1}{r}\,t_1^{-1/r}\int_0^{t_1}u^{-1+1/r}\,\e^{-u-gu^2}\,L_2(u,gu^2)\,\d u.
\eeq
This is an integral equation for the function $L_g(t_1)=L_2(t_1,gt_1^2)$,
which therefore obeys the differential equation
\beq
L_g'(t_1)=-\frac{1-\e^{-t_1-gt_1^2}}{rt_1}\,L_g(t_1).
\eeq
We thus obtain the explicit expression
\beq
L_2(t_1,t_2)=\e^{-F_2(t_1,t_2)/r},
\label{laps2}
\eeq
with
\beqa
F_2(t_1,t_2)&=&\int_0^1(1-\e^{-t_1x-t_2x^2})\frac{\d x}{x}
\nonumber\\
&=&\sum_{(m,n)\ne(0,0)}\frac{(-1)^{m+n-1}t_1^mt_2^n}{(m+2n)m!n!}.
\label{F2def}
\eeqa
The above result has the striking feature that
the generating function $F_2(t_1,t_2)$ obeys the inverted heat equation
\beq
\frac{\partial F_2}{\partial t_2}+\frac{\partial^2 F_2}{\partial t_1^2}=0,
\eeq
where $t_2$ plays the role of a negative time,
and with the initial condition $F_2(t_1,0)=F_1(t_1)$.
The series in the rightmost side of~(\ref{F2def})
demonstrates that the joint cumulants of $S_1$ and $S_2$ have the simple expression
\beq
\cum{S_1^mS_2^n}=\frac{1}{(m+2n)r}.
\eeq
The joint distribution of~$S_1$ and~$S_2$,
\beq
f_{S_1,S_2}(S_1,S_2)
=\int\!\!\!\int\frac{\d t_1}{2\pi\ii}\,\frac{\d t_2}{2\pi\ii}\,
\e^{t_1S_1+t_2S_2-F_2(t_1,t_2)/r},
\label{fcon2}
\eeq
is however highly non-trivial.

\subsection{Distribution of the quantity $Y$}

\begin{figure}[!ht]
\begin{center}
\includegraphics[angle=-90,width=.5\linewidth]{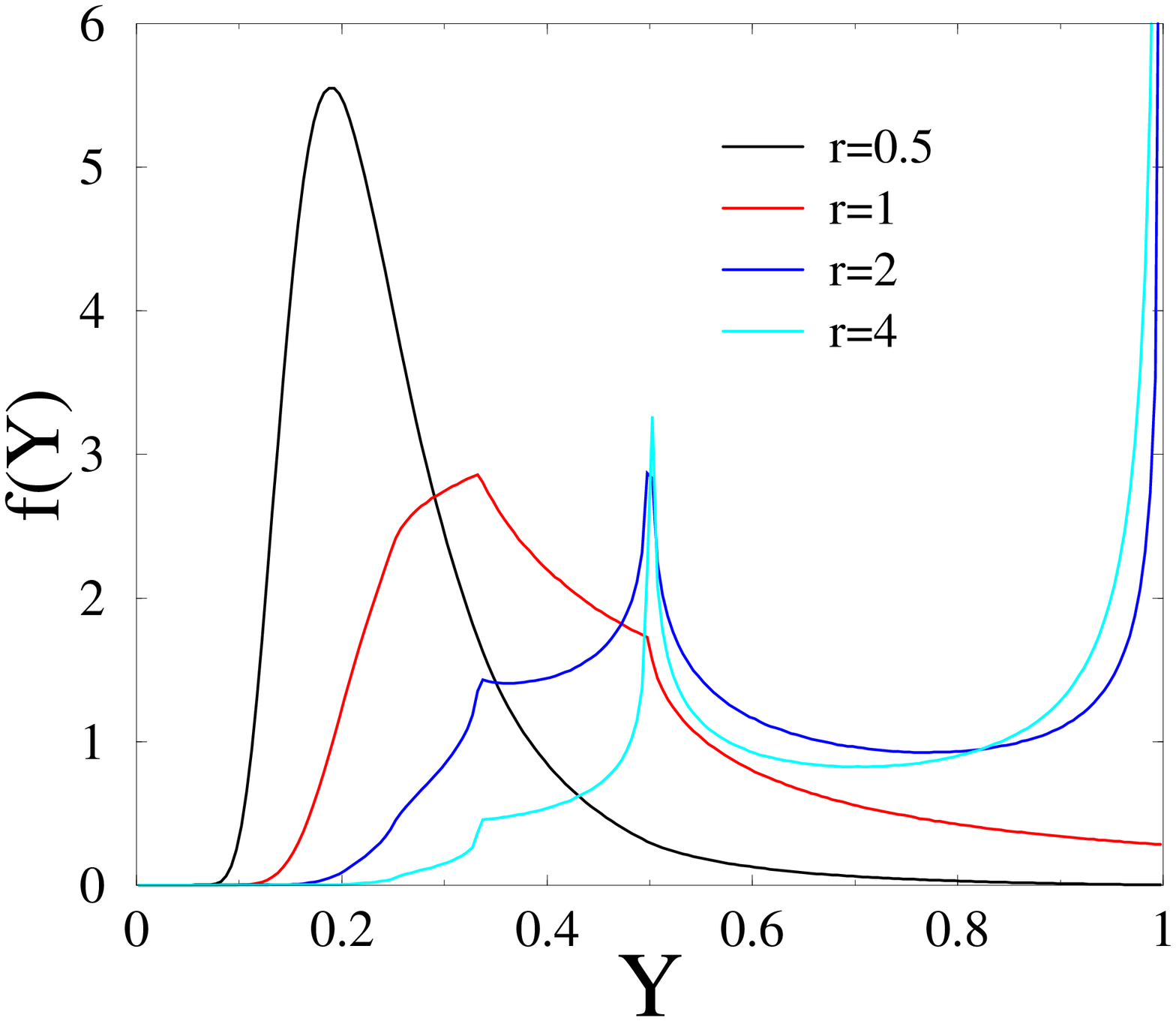}
\caption{\small
Distribution $f_Y(Y)$ of the quantity $Y$,
for four characteristic values of the control parameter $r$.}
\label{yhisto}
\end{center}
\end{figure}

The distribution of the quantity $Y$ introduced in~(\ref{ydef}) reads formally
\beq
f_Y(Y)=\int_0^\infty f_{S_1,S_2}(S_1,YS_1^2)\,S_1^2\,\d S_1,
\eeq
where the joint distribution $f_{S_1,S_2}(S_1,S_2)$ of the sums $S_1$ and $S_2$
is given by~(\ref{fcon2}).
The resulting triple integral formula is not very useful.
Figure~\ref{yhisto} shows a plot of the distribution $f_Y(Y)$
for four characteristic values of the control parameter $r$.
This distribution inherits from the joint distribution $f_{S_1,S_2}(S_1,S_2)$
a highly non-trivial structure,
with weaker and weaker singularities at the reciprocal integers
$Y=1/2$, $Y=1/3$, and so on.
As the control parameter $r$ gets larger and larger,
the overall distribution shifts towards larger values of $Y$,
and its singularities at $Y=1$ and $Y=1/2$ become more and more pronounced.

\begin{figure}[!ht]
\begin{center}
\includegraphics[angle=-90,width=.425\linewidth]{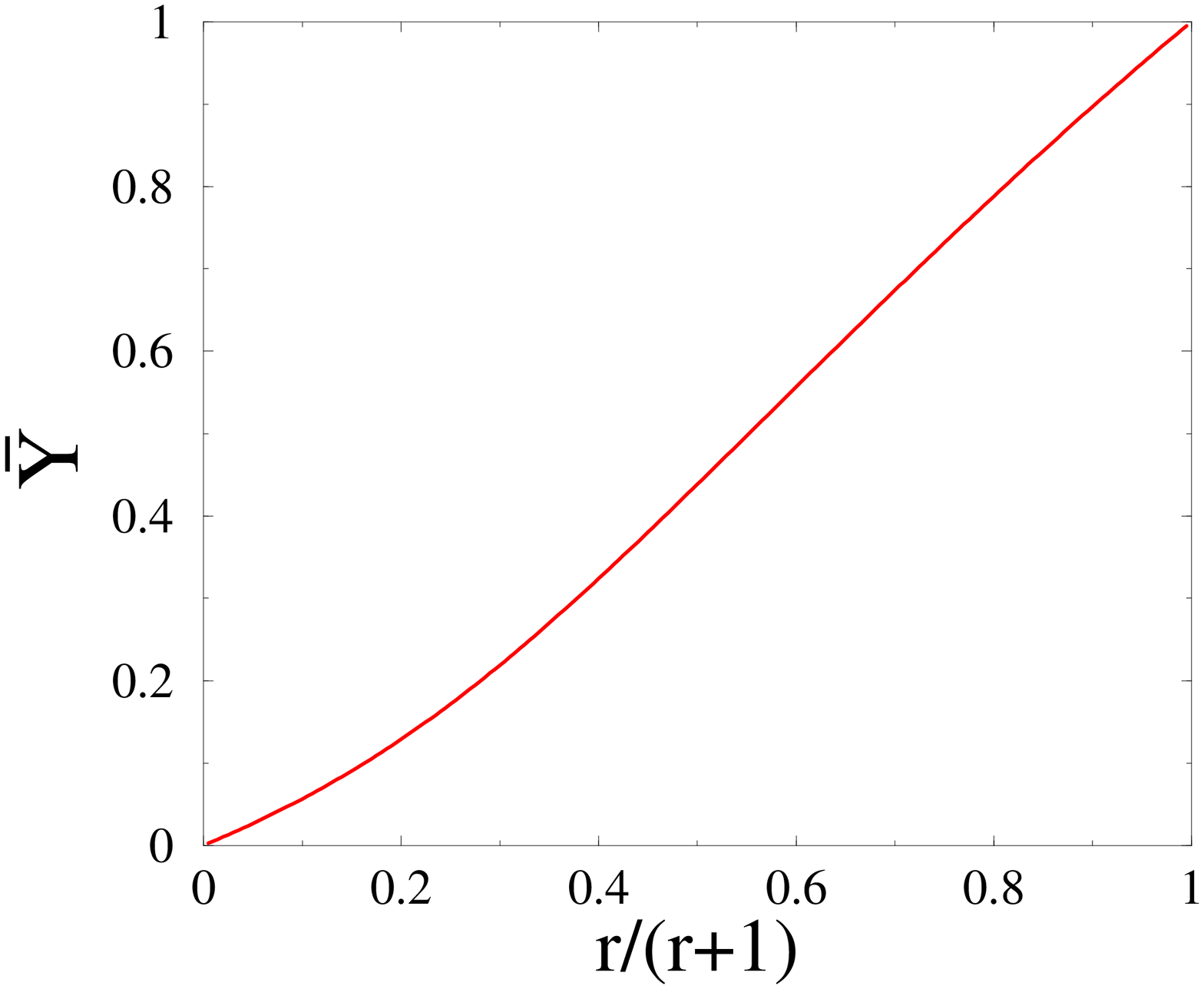}
{\hskip 8pt}
\includegraphics[angle=-90,width=.425\linewidth]{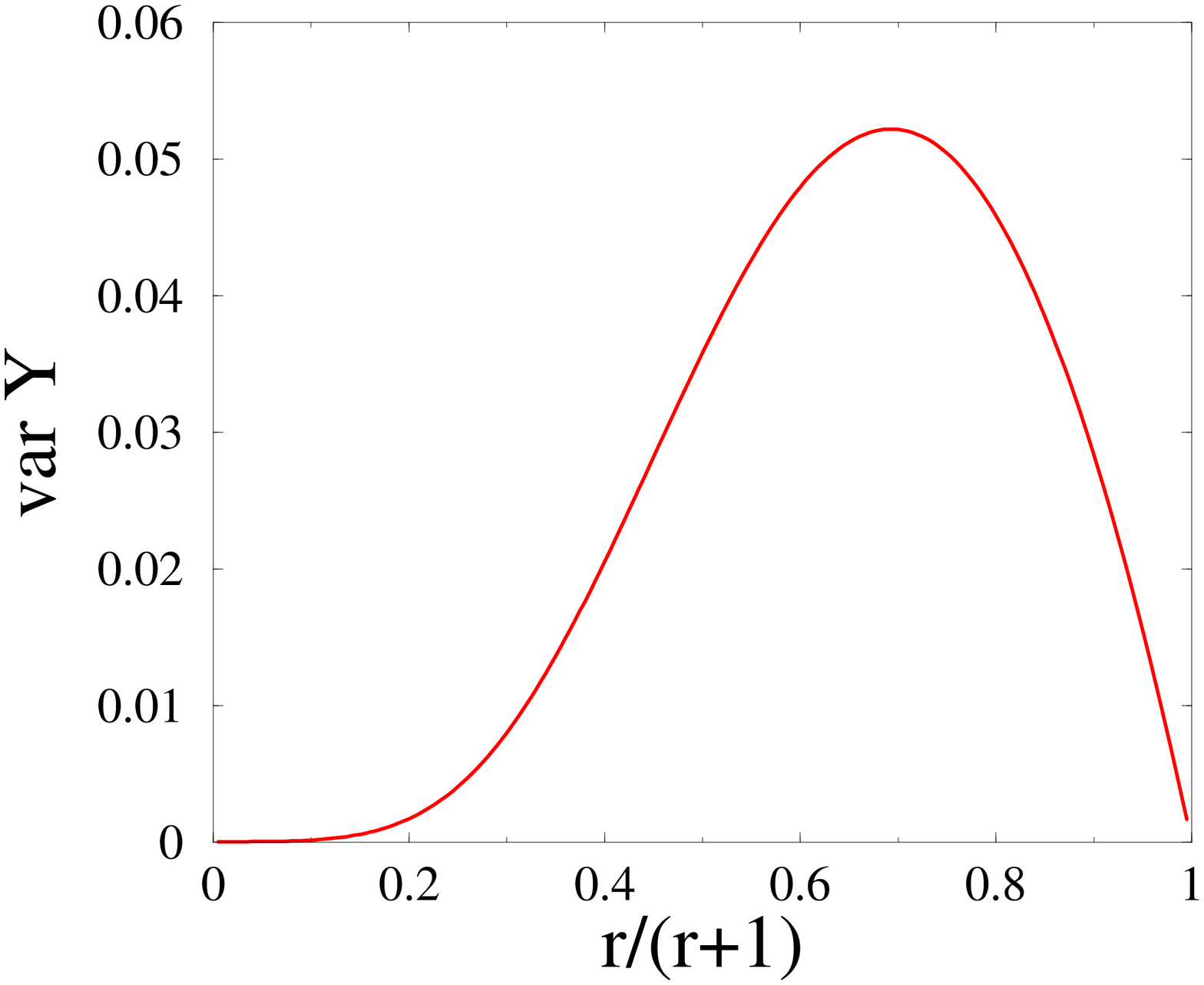}
\caption{\small
Plots of $\overline{Y}$ (left) and of $\var{Y}$ (right), against $r/(r+1)$.}
\label{ymom}
\end{center}
\end{figure}

The moments $\overline{Y^k}$ can be evaluated,
at least in principle, by means of the identity
\beq
Y^k=\frac{S_2^k}{S_1^{2k}}=\frac{S_2^k}{(2k-1)!}\int_0^\infty\e^{-tS_1}\,t^{2k-1}\,\d t.
\eeq
The quantities $\overline{S_2^k\e^{-tS_1}}$
can be calculated by expanding the result~(\ref{laps2}) as a Taylor series in $t_2$.
Some algebra involving integrations by parts
leads to the following integral expressions
for the first two moments of $Y$:
\beqa
\;\overline{Y}=1&-&\frac{1}{r}\int_0^\infty\e^{-F_1(t)/r}\,\e^{-t}\,\d
t,\nonumber\\
\overline{Y^2}=1&-&\frac{1}{6r}\int_0^\infty\e^{-F_1(t)/r}\,\e^{-t}(t^2+3t+5)\d
t\nonumber\\
&-&\frac{1}{6r^2}\int_0^\infty\e^{-F_1(t)/r}\,\e^{-t}(2-(t+2)\e^{-t})\d t.
\label{ymomint}
\eeqa
Figure~\ref{ymom} shows plots of $\overline{Y}$ (left)
and of $\var{Y}$ (right),
obtained by evaluating the above integrals numerically, against $r/(r+1)$.
The variance reaches its maximum $\var{Y}\approx0.05218$ for $r\approx2.25$.

The mean value of the quantity $R=1/Y$
can be determined along the same line of thought.
Some algebra using the identity
\beq
R=\frac{1}{Y}=\frac{S_1^2}{S_2}=S_1^2\int_0^\infty\e^{-tS_2}\,\d t
\eeq
leads to the following integral expression
\beq
\overline{R}
=1+\frac{\pi}{4r^2}\int_0^\infty\e^{-F_1(t)/(2r)}(\erf\sqrt{t})^2\,\frac{\d t}{t}.
\label{kapp}
\eeq

In the $r\ll1$ regime, where the number of significant weights gets large, we obtain
\beqa
\;\;\overline{Y}&=&\frac{r}{2}+\frac{r^2}{12}-\frac{r^3}{24}-\frac{101r^4}{720}+\cdots,
\nonumber\\
\var{Y}&=&\frac{r^3}{2}+\frac{37r^4}{360}+\frac{7r^5}{90}+\cdots,
\nonumber\\
\;\;\overline{R}&=&\frac{2}{r}+\frac{1}{3}+\frac{14r}{15}+\cdots,
\eeqa
and so the distribution of $Y$ is asymptotically a narrow Gaussian around $r/2$,
whose width scales as $r^{3/2}$.

In the opposite $r\gg1$ regime,
the distribution of $Y$ is dominated by a strong divergence as $Y\to1$,
and so the moments approach unity as
\beqa
\;\;\overline{Y}&=&1-\frac{1}{r}+\frac{\ln 2}{r^2}+\cdots,
\nonumber\\
\var{Y}&=&\frac{1}{3r}-\frac{4\ln 2+7}{12r^2}+\cdots,
\nonumber\\
\;\;\overline{R}&=&1+\frac{\pi}{2r}+\frac{\pi\ln 2-2\C}{2r^2}+\cdots,
\label{momto1}
\eeqa
where $\C=0.915\,965\dots$ denotes Catalan's constant.
This regime is studied in more detail in the next section.

\subsection{Effective two-site model in the $r\gg1$ regime}

In the $r\gg1$ regime,
the (ordered) weights $\Pi_m=\e^{-rx_m}$ fall off very rapidly with~$m$.
We are therefore led to consider an effective two-site model,
where only the largest two weights are kept,
while all the other ones are neglected.
The corresponding Poissonian points are $x_1=\tau_1$ and $x_2=\tau_1+\tau_2$.

Within this framework, we have
\beqa
&&S_1=\Pi_1+\Pi_2=\e^{-r\tau_1}(1+\e^{-r\tau_2}),\nonumber\\
&&S_2=\Pi_1^2+\Pi_2^2=\e^{-2r\tau_1}(1+\e^{-2r\tau_2}),
\eeqa
and hence
\beq
Y=\frac{1+\e^{-2r\tau_2}}{(1+\e^{-r\tau_2})^2}.
\eeq
The quantity $Y$ therefore only depends on $\tau_2$.
It lives in the interval $1/2<Y<1$, and its distribution reads
\beq
f_Y(Y)=\frac{1}{r(1-Y)\sqrt{2Y-1}}\left(\frac{1-Y}{Y+\sqrt{2Y-1}}\right)^{1/r}.
\eeq
This distribution exhibits an inverse-square-root singularity at the lower
edge:
\beq
f_Y(Y)\approx\frac{2}{r\sqrt{2Y-1}}\qquad(Y\to1/2),
\eeq
and a very strong divergence at the upper edge:
\beq
f_Y(Y)\approx\frac{1}{r}(1-Y)^{-1+1/r}\qquad(Y\to1).
\eeq

As a consequence, all the moments of $Y$ approach unity according to
\beq
\overline{Y^k}=1-\frac{\mu_k}{r},
\eeq
where the amplitude of the negative $1/r$ correction term reads
\beq
\mu_k=\int_{1/2}^1\frac{\d Y}{\sqrt{2Y-1}}\,\frac{1-Y^k}{1-Y}.
\label{mukdef}
\eeq
The generating series of these numbers can be evaluated as
\beqa
M(z)=\sum_{k\ge1}\mu_kz^k
&=&\frac{z}{1-z}\int_{1/2}^1\frac{\d Y}{\sqrt{2Y-1}(1-zY)}\nonumber\\
&=&\frac{1}{1-z}\sqrt\frac{z}{2-z}\ln\frac{1+\sqrt{z(2-z)}}{1-z}.
\eeqa
This expression yields the values $\mu_1=1$ and $\mu_2=5/3$
(in agreement with~(\ref{momto1}) to order $1/r$),
$\mu_3=32/15$, $\mu_4=52/21$, $\mu_5=863/315$, and so on,
as well as the linear recursion formula
\beq
(2k-1)\mu_k-(3k-2)\mu_{k-1}+(k-1)\mu_{k-2}=1,
\eeq
and the asymptotic growth law $\mu_k\approx\ln(2k)+\gamma$.

Finally, the formal expression obtained by setting $k=-1$ in~(\ref{mukdef}),
i.e.,
\beq
\mu_{-1}=-\int_{1/2}^1\frac{\d Y}{Y\sqrt{2Y-1}}=-\frac{\pi}{2},
\eeq
agrees with the expansion~(\ref{momto1}) for $\overline{R}$.

\section{Effective two-site model in the localized regime ($c>1$)}
\label{effloc}

The effective model studied in~\ref{borderline}
can be extended to the localized regime ($c>1$),
in order to describe the rare events
where the first two weights $\Pi_1$ and $\Pi_2$ are comparable.
We still have $S_1=\Pi_1+\Pi_2$ and $S_2=\Pi_1^2+\Pi_2^2$, and so
\beq
Y=\frac{1+\eta^2}{(1+\eta)^2},\qquad
R=\frac{(1+\eta)^2}{1+\eta^2}=1+\frac{2\eta}{1+\eta^2},
\label{yketa}
\eeq
where the weight ratio $\eta=\Pi_2/\Pi_1$ reads~(see~(\ref{pisca}))
\beq
\eta\approx\exp\left(-\frac{\Delta}{M^{1/c}}\left(x_2^{1/c}-x_1^{1/c}\right)\right).
\eeq

Let us first evaluate the mean value of this ratio:
\beq
\overline{\eta}\approx\int_0^\infty\e^{-x_2}\d x_2\int_0^{x_2}
\exp\left(-\frac{\Delta}{M^{1/c}}\left(x_2^{1/c}-x_1^{1/c}\right)\right)\d x_1.
\eeq
The integral over $x_1$ is dominated by small values of the difference
$\tau_2=x_2-x_1\sim M^{1/c}/\Delta\sim M^{-(1-1/c)}\ll1$.
Linearizing the argument of the exponential with respect to $\tau_2$,
and integrating over $\tau_2$ first, we readily obtain
\beq
\overline{\eta}\approx c\,\Gamma(2-1/c)\,\frac{M^{1/c}}{\Delta}
=\frac{c\,\Gamma(2-1/c)}{\rho-\rho_c}\,M^{-(1-1/c)}.
\eeq
Higher moments of $\eta$ can be estimated
by replacing in the above result $\Delta$ by the product $k\Delta$.
We thus get immediately
\beq
\overline{\eta^k}\approx\frac{\overline{\eta}}{k}.
\label{etamom}
\eeq
This result leads to the picture that $\eta$ is zero with very high probability,
and of order unity with a small probability of order $\overline{\eta}$.
The quantity $\overline{\eta}$ thus provides an operational estimate for the
probability that $\Pi_1$ and $\Pi_2$ are comparable.
This probability was already anticipated in section~\ref{thermo}
to fall off as $M^{-(1-1/c)}$.

Finally, by expanding the expression~(\ref{yketa}) for $R$ as a power series
in $\eta$, using~(\ref{etamom}),
as well as the result\footnote{This identity is attributed to James Gregory
(1638--1675).}
\beq
1-\frac13+\frac15-\frac17+\cdots=\frac{\pi}{4},
\eeq
we are left with the following quantitative estimate for the mean value of $R$
in the localized regime:
\beq
\overline{R}-1\approx\frac{\pi c\,\Gamma(2-1/c)}{2}\,\frac{M^{1/c}}{\Delta}
=\frac{\pi c\,\Gamma(2-1/c)}{2(\rho-\rho_c)}\,M^{-(1-1/c)}.
\label{appkloc}
\eeq

\end{document}